\documentstyle[12pt]{article}
\newcommand{\be}{\begin{equation}}
\newcommand{\ee}{\end{equation}}
\newcommand{\lb}[1]{\label{#1}}
\font \greekb=cmmib10  scaled \magstep1
\font \greekbi=cmmib7  scaled 1440 %\magstep1
\font \gothic=eufm10  scaled \magstep1
\font \mathbf=cmbxti10  scaled \magstep1
\font \mathbfi=cmbxti7  scaled 1440 %\magstep1
\newcommand\mbf[1]{{\mbox{{\mathbf #1}}}}
\newcommand\mbfi[1]{{\mbox{{\mathbfi #1}}}}
\newcommand\goth[1]{{\mbox{{\gothic #1}}}}
\newcommand{\nub}{\mbox{\greekb \char 23}}
\newcommand{\nubi}{\mbox{\greekbi \char 23}}
\begin{document}
\pagenumbering{arabic}
\thispagestyle{empty}
\noindent
\phantom{--------------------------------------------------------------------
------------------}NPI MSU -- 99-10/568\\[10mm]
\begin{center}
\begin{Large}
{\bf Finslerian Spaces\\[2mm]
 Possessing Local Relativistic Symmetry}\\[20mm]
\end{Large}
{\large G.Yu.~Bogoslovsky\\
{\it Institute of Nuclear Physics, Moscow State University\\
119899, Moscow, Russia}\\
E-mail: bogoslov@theory.npi.msu.su\\
\vspace{4mm}
 H.F.~Goenner\\
{\it Institute for Theoretical Physics, University of G\"ottingen\\
Bunsenstr.9, D-37073, G\"ottingen, Germany}\\
E-mail: goenner@theorie.physik.uni-goettingen.de}\\[20mm]
\end{center}

It is shown that the problem of a possible violation of the Lorentz
transformations at Lorentz factors $\gamma >5\times 10^{10}\,,$ indicated
by the situation which has developed in
the physics of ultra-high energy cosmic rays (the absence of the GZK cutoff),
has a nontrivial solution. Its essence consists in the discovery of the
so-called generalized Lorentz transformations which seem to correctly link the
inertial reference frames at any values of $\gamma \,.$ Like the usual Lorentz
transformations, the generalized ones are linear, possess group properties and
lead to the Einstein law of addition of 3-velocities. However, their geometric
meaning turns out to be different: they serve as relativistic symmetry
transformations of a flat anisotropic Finslerian event space
rather than of Minkowski space. Consideration is given to two types of
Finsler spaces which generalize locally isotropic Riemannian
space-time of relativity theory, e.\,g. Finsler spaces with a
partially and entirely broken local 3D isotropy. The investigation
advances arguments for the corresponding generalization of the
theory of fundamental interactions and for a specific search for
physical effects due to local anisotropy of space-time.\\[10mm]
\noindent
24/03/99\\
\newpage
\noindent
{\bf{1. INTRODUCTION}}

\bigskip
\bigskip

\noindent
At present, apart from general relativity theory (GR),
there exist a number of alternative metric theories of gravitation. They all 
employ the Riemannian geometric model of space-time borrowed from GR,
and differ only by the field equations which describe the
self-consistent dynamics of space-time and matter. The cosmological
models based on such theories differ accordingly. Common to them,
however, is the fact that space-time, being Riemannian and,
consequently, locally isotropic, preserves its local isotropy during
the evolution of the Universe. 

Although, as it would seem, there is no reason to question the local isotropy
of space (the more so as no violation of the law of angular momentum
conservation has yet been revealed), there are some indirect indications
that in our epoch space-time, on the average, has a weak relic local
anisotropy, and that it therefore should be described by Finsler
geometry [1] rather than by Riemann geometry. A strong local
anisotropy of space-time might have occured at an early stage in the 
evolution of the Universe as a result of high-temperature phase
transitions in its geometric structure, caused by a
breaking of higher gauge symmetries and by the appearance of massive 
elementary particles. If this was the case, it is natural to assume
that the local anisotropy of space decreased to its present low
level $( < 10^{-10})$ due to the expansion of the Universe. 

The existence of a local anisotropy of space-time is indirectly indicated by
the following facts: (i) a breaking  of the discrete space-time symmetries in
weak interactions; (ii) an anisotropy of the relic background radiation
filling the Universe; and (iii) the absence [2] of the effect of cutoff of the
spectrum of primary ultra-high energy cosmic protons, i.e.\ of the so-called GZK
cutoff [3,4]. 

By a strict local isotropy of Riemannian space-time we imply that, at
each point, its tangent space is Minkowski space\footnote{ The name Minkowski
space is
used here in
the usual sense, i.e. for a 4-dimensional, pseudo-Euclidean, flat
Riemannian space. In contrast, Rund [1] has used it for a flat Finsler
space.}, the isotropic event
space of special relativity theory (SR). In Galilean coordinates, the
pseudo-Euclidean metric is of the form $ds^2=dx_0^2-d\mbf x^{\,2}\,.$ 
As under the discrete transformations: $x_0\to -x_0, x_\alpha
\to -x_\alpha ,$ this metric is invariant under the continuous 
transformations belonging to the 10-parameter inhomogeneous Lorentz or
Poincar\'e group.\footnote{Four parameters correspond to space-time 
translations, three to 3D rotations, and another three to Lorentz boosts.} 

From the mathematical point of view, the presence of the Poincar\'e
group as a group of relativistic symmetry (isometry group) of the
event space is the necessary and sufficient condition for it to be
Minkowski space. Therefore, if the Poincar\'e symmetry turns out to be only
approximate, and if the exact transformations of relativistic symmetry
realized in nature are some ``generalized Lorentz transformations''
imbedded into another group, then the event space has a geometry
different from that of Minkowski space - even at the level of SR. 

The idea of a possible violation of the usual Lorentz transformations
at Lorentz factors $\gamma >5\times 10^{10} $, and of a corresponding 
generalization of the relativistic theories was suggested first in
[5,6]. Its motivation rested on a discrepancy, assumed at the time,
between the theoretical predictions [3,4] and the experimental data
[7] relating to the behaviour of the spectrum of primary ultra-high
energy cosmic protons. If the usual Lorentz transformations would
correctly link inertial frames at relative velocities very close
to the velocity of light, then, in the case of uniformly distributed
sources, the energy spectrum of primary cosmic protons should show a
cutoff (due to inelastic collisions of the protons with cosmic
background radiation photons) at proton energies $\sim 5\times 10^{19}
eV.$  However, as now has been firmly established, such a prediction
is at variance with present experimental data\footnote{In connection
  with this situation S. Coleman and S. L. Glashow [8] argue that 
possible departures from strict Lorentz invariance can affect 
elementary-particle kinematics so as to suppress or forbid inelastic collisions
of cosmic-ray nucleons with background photons.}. 

Apart from the violation of the Lorentz transformations, there exist
also other possible causes of the absence of the GZK cutoff
[9]. Nevertheless, the assumption that the inertial frames could be
linked by some ``generalized Lorentz transformations'' markedly
different from the usual Lorentz transformations only at relative velocities
extremely close to the velocity of light, remains valid. Moreover,
general considerations make it possible to find the required
transformations in an explicit form. There exists an 8-parameter group
of relativistic symmetry obviously different from the Poincar\`e group
[10]. Along with space-time translations and the ``generalized Lorentz
transformations'' (three parameters), the group includes only a
1-parameter subgroup of rotations of 3D space about some preferred
direction. Since, as it turned out, such an 8-parameter group allows
for a geometric invariant in the form of a flat Finsler metric
generalizing the Minkowski metric of SR, the door is opened to a nontrivial 
generalization of relativity theory [11-13]. 

Although any relativistic theory is constructed from the requirement of 
invariance of its equations under the Poincar\'e group, soon after the
creation of SR, the authors of [14,15] demonstrated invariance of the 
electrodynamic equations not only under Poincar\'e group but also
under the 15-parameter conformal group. This group incorporates both linear
and nonlinear transformations of event coordinates [16]. In terms of SR, the
nonlinear transformations are of no interest since they link noninertial
frames. A full classification of the subgroups of linear transformations of
the conformal group has not yet been carried out. From the very
outset, however, it was known that one of the linear subgroups of the 
conformal group is the Poincar\'e group. 

In the next section it will be shown that the above-mentioned
8-parameter group\footnote{i.e. the group whose invariant is the
  Finslerian metric describing a flat anisotropic event space} is
another linear subgroup of the conformal group. It likewise leads to 
Einstein's law of addition of 3-velocities. Therefore, if
relativistic physics could have been developed already during a
cosmological epoch with a sufficiently large local anisotropy of
space and a flagrant violation of the conservation law of total
angular momentum for any closed physical system, then the 8-parameter
linear subgroup of the conformal group rather than the Poincar\'e
subgroup might have been chosen as a group of relativistic
symmetry. In order to empirically test such a choice in our
epoch, very specific experiments are required since 
the rate of change of the total angular momentum is the lower, the smaller
the magnitude of local space-time anisotropy will be. This signifies that 
non-conservation of angular momentum may primarily be manifested in 
processes for which the interaction time is long enough. 

When speaking of a flat locally anisotropic event space, we implied a space
which, while not being symmetric with respect to arbitrary 3D
rotations, is still symmetric with respect to rotations around some 
preferred direction and therefore will be referred to as a space with
a partially broken 3D isotropy. Although as one of the consequences of
this breaking of symmetry a violation of the conservation law of total
angular momentum results, this is not inconsistent, but permissible from
the point of view of physics. A demonstration is given by the analysis
[17] of the corresponding generalized Dirac equation with the existence of
stable massive fermions. Of physically interest seems also a model of 
space-time [18,19] with an {\em entirely} broken local isotropy
corresponding to a 7-parameter inhomogeneous group of relativistic symmetry. 

In  section 2, we present a model of a flat Finsler space with
a partially broken local rotational symmetry and sketch the induced
point mechanics. Section 3 introduces a flat Finsler space with a
totally broken local rotational symmetry. Here, the second simplest type of
line elements for Finsler spaces already mentioned by Riemann [20]
appears. Properties of the isometry group and some physical
consequences of this model are discussed in sections 4 and 5. In both
models the conformal structure of SR is preserved. Finally,
section 6 briefly reviews a generalization to curved Finsler space,
eventually needed for an interpretation of recent observational discoveries in
astrophysics.

\bigskip
\bigskip

\noindent
{\bf 2.\,\, A RELATIVISTICALLY SYMMETRIC FINSLERIAN SPACE-TIME\\ 
\phantom{AA}WITH  PARTIALLY BROKEN 3D ISOTROPY

\bigskip
\bigskip
\noindent
2.1 Introducing the model}

\bigskip

In order to arrive at a viable Finslerian model of space-time, for the
sake of simplicity we first confine ourselves to a two-dimensional
space and show that it is possible to generalize the Lorentz transformations 
\be \lb{1}
\left\{
\begin{array}{lll}
{x_0^{\prime}}&=&x_0\cosh \alpha -x\sinh \alpha \\ 
x'            &=-&x_0\sinh \alpha +x\cosh \alpha \,;
                       \qquad \tanh \alpha =v/c\\
\end{array}
\right.
\ee 
so that the new linear transformations will also form a group with
a single parameter $\alpha$ and will keep invariance of the wave
equation $(\partial^2/\partial x_0^2-\partial^2/\partial x^2)\,f=0\,.$ 
Guided by the conformal invariance of the electrodynamic
equations, we insert an additional scale transformation into (1). As a result,
in place of (1), we obtain the generalized Lorentz transformations in the form
\be\lb{2}
\left\{
\begin{array}{lclll}
   x'_0&=&e^{-r\alpha}&(&x_0\cosh\alpha -x\sinh\alpha)\\
   x'&=&e^{-r\alpha}&(-&x_0\sinh\alpha +x\cosh\alpha)\,,
\end{array}
\right.
\ee 
where $r$ is a dimensionless parameter of the scale transformation. Since
according to (2) the relation of the group parameter $\alpha$ to the velocity  
$v$ of the primed frame remains the same, i.e.\ $\tanh\alpha =v/c$\,,
(2) can be rewritten as follows
\be\lb{3}
\left\{
\begin{array}{lcl}
x'_0&=&\left(\frac{1-v/c}{1+v/c}\right)^{r/2}
\frac{x_0-(v/c)x}{\sqrt{1-v^2/c^2}}\\
x'&=&\left(\frac{1-v/c}{1+v/c}\right)^{r/2}
\frac{x-(v/c)x_0}{\sqrt{1-v^2/c^2}}\,.
\end{array}
\right.
\ee

Obviously, in contrast to (1), the ``generalized Lorentz
transformations'' (2) or (3) do not leave invariant the pseudo-Euclidean metric
$ds^2=dx_0^2-dx^2$ but conformally modify it. Therefore, the
question arises as to what the metric of an event space invariant
under such ``generalized Lorentz transformations'' is. The rigorous
solution to this problem is 
\be\lb{4}
ds^2=\left[\frac{(dx_0-dx)^2}{dx_0^2-dx^2}\right]^r(dx_0^2-dx^2)\,.
\ee    
Not being a quadratic form but a homogeneous function of the coordinate
differentials of degree two, the metric (4) falls into the category of
Finsler metrics. It describes a flat but anisotropic event
space.\footnote{In accordance with the Busemann approach to the mathematical
theory of Finsler spaces, a flat anisotropic space is understood as a
linear normalized vector space in which the norm of a vector is
determined not exclusively by its Euclidean length but also by its
orientation with respect to some preferred direction.}
As long as we deal with 2D anisotropic space, its 
anisotropy manifests itself in the noninvariance of the metric (4) under the 
reflections $x_0\to -x_0$ or $x\to -x\,.$ If $r=0\,,$ then the
anisotropy disappears. In this case, the event space becomes isotropic
while the ``generalized Lorentz transformations'' (3) reduce to the
usual Lorentz transformations. However, if $r\ne 0$
characterizing the magnitude of space anisotropy, is sufficiently small,
then the additional dilatation of space-time, which distinguishes the
generalized Lorentz transformations from the usual ones, becomes markedly
different from unity only at relative velocities of the inertial 
frames extremely close to the velocity  of light. In the physics of 
ultra-high energy cosmic rays we deal with precisely such a
situation. Therefore, the use of the ``generalized Lorentz
transformations'' instead of the usual ones makes it possible, in
principle, to remove the discrepancy between theory and experiment in
this field; this may be regarded as a hint towards a local anisotropy
of space. 

According to (4), the parameter $r$ is limited by the condition $|r|<1$. 
In this case, due to equation $ds^2 = 0\,,$ the velocity of light is
the same in both directions of the $x$- axis and is equal to $c$ in
spite of the presence of a space-time anisotropy. Both
relations (3) incorporate the same power-type scale factor canceling
out when the second relation (3) is divided by the first one. As a
result the Einstein formula, which describes the addition of 3D
velocities, is reproduced: 
\be\lb{5}
V=\frac{V'+v}{1+V'v/c^2}\,\, ,
\ee
where $v$ is the velocity of the primed frame. 

Certainly, the 2D model (4) of a flat anisotropic event space is of
methodical interest only and must be generalized to the 4D case. It
turns out that two independent ways for such an extension exist. The
first path leads to a flat anisotropic space-time with
partially broken symmetry with respect to 3D rotations. The second way
ends in a flat anisotropic space-time with entirely broken rotational symmetry.
Both models are sufficiently interesting from a physical aspect. 

\bigskip
\bigskip

\noindent
{\bf 2.2 Partially broken isotropy}

\bigskip

We first will discuss the model of space-time with a partially broken 
isotropy. The corresponding 4D Finslerian metric can be found if we
note that the 2D metric (4) is a definite function of two
quantities: $(dx_0^2-dx^2)$ and $(dx_0-dx)\,.$ The first quantity is
the pseudo-Euclidean square of a
vector having components $(dx_0,dx)\,.$ The second quantity is the
pseudo-Euclidean scalar product of such a vector with a vector 
$\nu ^i=(1,1)\,.$ Replacing now the cited vectors by their 4D analogs with the
components $(dx_0,d\mbf x)$ and $\nu ^i=(1,\nub )\,,$ where 
$\nub ^{\,2}=1\,,$ we arrive at  
$$
(dx_0^2-dx^2)\rightarrow (dx_0^2-d\mbf x^{\,2})\,;\quad 
(dx_0-dx)\rightarrow (dx_0-\nub d\mbf x)\,,
$$
by means of which the 4D metric is obtained from the 2D metric (4)
\be\lb{6}
 ds^2=\left[\frac{(dx_0-\nub d\mbf x)^2}{dx_0^2-d\mbf x^{\,2}}\right]^r
(dx_0^2-d\mbf x^{\,2})\,.
\ee
This Finslerian metric depends on two constant parameters $r$ and $\nub$
and describes a flat anisotropic space-time with partially broken rotational
symmetry. Instead of the 3-parameter group of rotations of Minkowski space, the
space-time (6) admits only the 1-parameter group of rotations about
the unit vector $\nub\,,$ which indicates a preferred
direction in 3D space. No changes occur for translational symmetry: 
space-time translations leave the metric (6) invariant. As regards the
transformations linking the various inertial frames, the usual Lorentz
boosts modify the metric (6). Therefore, they do not belong to the
isometry group of the space-time (6). By proper use of them, however, 
invariance transformations for the metric (6) can be constructed. The
corresponding transformations, named ``generalized Lorentz
transformations'', will be the following
\be\lb{7}
x'^i=D(\mbf v,\nub )\,R^i_j(\mbf v,\nub )\,L^j_k(\mbf v)\,x^k\,.
\ee
In order to demonstrate how the invariance of the metric (6) is realized under
the transformations (7), we first carry out the transformation using the matrix
$L^j_k(\mbf v)\,,$ which represents a usual Lorentz boost (the given
matrix naturally depends on the velocity $\mbf v$ of a moving
frame). As a result, in expression (6) only the Minkowskian scalar
product $dx^0-\nub d\mbf x\,$ will change its form since the
components $(1,\nub )$ of the 4-vector $\nu^i$ will change. The vector $\nu^i$
was initially defined as light-like $(1-\nub ^{\,2}=0)\,.$
 It will remain the same after the boost, i.e.\ it will remain light-like
 although both the time
and spatial components are changed by a scale factor. In addition, the
spatial component of the 4-vector $\nu ^i$ will change its direction due to the
rotation about the vector $[\mbf v\,\nub ]$ through an angle of relativistic 
aberration
\be\lb{8}
\varphi=\arccos\left\{ 1-\frac{(1-\sqrt{1-\mbf v^{\,2}/c^2})
[\mbf v\,\nub ]^2}{
(1-\mbf v\nub /c)\mbf v^{\,2}}\right\}\,.
\ee
Therefore, having carried out (according to (7)), subsequent to the Lorentz
boost, an additional rotation $R^i_j(\mbf v,\nub )$ of the space axes
of the moving frame through the angle (8) about the vector 
$[\mbf v\,\nub ]\,,$ we regain for the spatial component of the
4-vector $\nu ^i$ its initial orientation
with respect to the space axes. The net result of the transformations
performed will be the fact that the form $(dx_0^2-d\mbf x^{\,2})$ will
not be changed while the form $(dx_0-\nub d\mbf x)$ will be altered
minimally: its new representation will be distinguished from the
initial one only by a scale factor dependent on $\mbf v$ and $\nub\,.$ If now,
as prescribed by (7), we perform
also the dilatation transformation of the event coordinates using the matrix 
\be\lb{9}
D(\mbf v,\nub)=\left(\frac{1-\mbf v\nub /c}{\sqrt{1-\mbf v^{\,2}/c^2}}
\right)^rI\,,
\ee
where $I$ is the unit matrix, then within the square brackets in (6) no
additional scale factor will appear\footnote{since the ratio, enclosed
  within these brackets, is homogeneous of degree zero with respect to
  the  coordinate differentials}, while the scale factor, which occurred there
  at the
previous stage of transformations, will be completely cancelled out by a scale
factor arising from the dilatation transformation of the expression
$(dx_0^2-d\mbf x^{\,2})\,.$ The result is that the generalized Lorentz
transformations determined by (7) do indeed leave the metric (6) invariant. 

In contrast to Lorentz boosts, the generalized
transformations (7) make up a 3-parameter noncompact group with generators
$X_1\,, X_2\,, X_3\,$. Thus, with the inclusion of the 1-parameter group of
rotations about the preferred direction $\nub\,$ and of the
4-parameter group of translations, the inhomogeneous group of
isometries of the space (6) turns out to have 8-parameters. In order
to obtain the simplest representation for its
generators, it is sufficient to choose a third space axis along
$\nub\,$ and then to make use of the infinitesimal form of the transformations
(7). As a result, 
\be \lb{10}
\left.
\begin{array}{rcl}
X_1&=&-(x^1p_0+x^0p_1)-(x^1p_3-x^3p_1)\,,\\
X_2&=&-(x^2p_0+x^0p_2)+(x^3p_2-x^2p_3)\,,\\
X_3&=&-rx^ip_i-(x^3p_0+x^0p_3)\,,\\
R_3&=&x^2p_1-x^1p_2\,; \qquad \qquad \qquad \qquad \qquad p_i=\partial /\partial
x^i\,.\\
\end{array}
\right.
\ee
The generators (10) satisfy the commutation relations 
\be \lb{11}
\left.
\begin{array}{llll}
  [X_1X_2]=0\,, & [R_3X_3]=0\,, && \\
  \left[X_3X_1\right]=X_1 \,, & [R_3X_1]=X_2 \,, && \\
  \left[X_3X_2\right]=X_2\,, & [R_3X_2]=-X_1\,; & & \\
  \left[p_i p_j\right]=0\,; &&& \\
  \left[X_1p_0\right]=p_1\,,& [X_2p_0]=p_2\,, & [X_3p_0]=rp_0+p_3\,, &
  [R_3p_0]=0\,, \\
  \left[X_1p_1\right]=p_0+p_3\,,& [X_2p_1]=0\,, & [X_3p_1]=rp_1\,, &
  [R_3p_1]=p_2\,, \\
  \left[X_1p_2\right]=0\,, & [X_2p_2]=p_0+p_3\,, & [X_3p_2]=rp_2\,, &
  [R_3p_2]=-p_1\,, \\
  \left[X_1p_3\right]=-p_1\,, & [X_2p_3]=-p_2\,, & [X_3p_3]=rp_3+p_0\,, &
  [R_3p_3]=0\,. \\
\end{array}
\right.
\ee
From (11), we conclude in particular that the homogeneous isometry
group of the space (6) contains 4 parameters (the generators 
$X_1\,,\, X_2\,,\, X_3\,,\, R_3$ ). Being a
subgroup of the conformal group, it is isomorphic to the corresponding
4-parameter subgroup of the homogeneous Lorentz group (with the
generators $X_1\,,\, X_2\,,\, X_3\!\!\mid_{r=0}\,,\, R_3$ ). Since
the 6-parameter homogeneous Lorentz group has no 5-parameter subgroup
[21] while the 4-parameter subgroup is unique (up to isomorphisms),
the transition from Minkowski space to the event space (6) implies a
minimum of symmetry-breaking of the Lorentz
symmetry. Some types of Finslerian spaces with more radical breaking of the
Lorentz symmetry are considered in [22--24]. 

A remarkable property of the anisotropic event space (6) is the fact
that it keeps the conformal structure (light cones) of Minkowski space,
i.e.\ light propagates according to the equation $dx_0^2-d\mbf x^{\,2}=0\,.$
Therefore, the velocity of light is independent of the direction of its
propagation and is equal to $c\,.$ It thus appears that the square of the
distance $dl^2$ between adjacent points of 3D space, determined by
means of exchange of light signals\footnote{As regards the additional
tachyon solution $\,dx_0-\nubi d\mbfi{x}=0\,$ of the equation
$ds^2=0\,,$ it does not admit the reflection operation $\,d\mbfi{x}\to -d\mbfi
x\,.$ Therefore, there is no algorithm
for determining the 3D distances based on exchange of tachyon signals.}\,,
is expressed by the formula $dl^2=d\mbf x^{\,2}\,.$ Thus, although in
the 3D space there is a preferred direction $\nub\,,$ its geometry
remains Euclidean. But, what does the anisotropy physically manifest
itself in? First of all, it affects the dependence of proper time of a
moving clock by including the {\em direction} of its velocity in
addition to the magnitude. According to (6), the interval $d\tau$ of
proper time read by the clock moving with a velocity $\mbf v\,,$
is related to the time interval $dt$ read by a clock at rest by the relation 
$d\tau =(d\tau /dt)\,dt\,,$ where 
\be\lb{12}
\frac{d\tau}{dt} =\left(\frac{1-\mbf v\nub /c}{
\sqrt{1-\mbf v^{\,2}/c^2}}\right)^r\sqrt{1-\mbf v^{\,2}/c^2}\,\,.
\ee
It can be seen from Fig.\,1\,, that, in contrast to Minkowski space (for which:
$r=0\,,$ $(d\tau /dt)\!\!\mid_{r=0}\;=\sqrt{1-v^2/c^2}\le 1\,$ and, hence, 
the moving clock is
always slow in comparison with the clock at rest), in the anisotropic space the
time dilatation factor $(d\tau /dt)\!\!\!\mid_{r>0}$ can take on values
greater than unity. Therefore, at some of its velocities the clock
moving in the anisotropic space is fast in comparison with the clock
at rest. However, having returned to its starting point, it will
necessarily run behind the clock at rest. Consequently, at $r>0$ inertial motion
is still
uniform and along a straight line. 
 
Along with the time dilatation factor the anisotropy of space also affects the
Doppler shift. In place of the usual relativistic formula, now the modified
relation [25] holds: 
\be \lb{13}
\omega =\omega '\,\frac{\sqrt{1-\mbf v^{\,2}/c^2}}{1-\mbf v\mbf e/c}
\left (\frac{1-\mbf v\nub /c}{\sqrt{1-\mbf v^{\,2}/c^2}}\right )^r\,,
\ee
where $r$ is the magnitude of space anisotropy, $\mbf v$  the velocity of a
moving frame, $\omega '$  the frequency of a ray with respect to it, and
$\omega\ ,\mbf e$ and $\nub $ are the frequency, direction of the ray
and the preferred direction in an initial frame. 

Precision measurements of the Doppler effect by use of the
M\"{o}ssbauer effect were suggested and have since been made
(Cf. [26--28]). Of special interest here is the experiment of ref. [27] which 
consisted in measuring a relative frequency shift $\Delta\,\omega
/\omega = (\omega _a-\omega _s)/\omega _s$ 
between a M\"{o}ssbauer source and an absorber placed at
equal and diametrically opposite distances from the center of a rapidly
rotating rotor. For the quantity $\Delta\,\omega /\omega $, the 
prerelativistic theory of absolute aether (PR), SR and the
relativistic theory of locally anisotropic space (AR), respectively, give the
following predictions to within $v^2/c^2$ 
\begin{eqnarray}
& &\left (\Delta\,\omega /\omega \right )^{PR} =2\mbf w\mbf v_a/c^2\,, \\
& &\left (\Delta\,\omega /\omega \right )^{SR} =  0\,, \\
& &\left (\Delta\,\omega /\omega \right )^{AR}  =  2rc\nub \mbf
v_a/c^2\,, 
\end{eqnarray}
where $\mbf w$ is the velocity of the aether wind and $\mbf v_a$  the 
velocity of the absorber. Comparing (14) and (16), we may regard the 
quantity $rc\nub $ in a sense as the velocity of the aether wind. It
must be noted, however, that $rc\nub $ is an invariant of the
generalized Lorentz transformations. In the experiment of ref. [27],
no aether wind was found. As a result, an upper limit, specified in
ref. [29], was obtained for the velocity of the aether wind. In terms of the
relativistic theory of anisotropic space-time this constraint signifies that
the value of anisotropy $r < 5\times 10^{-10}$. At the present time, due to the
use of radically new rotors developed at the I.T.E.P. (Moscow) and of
M\"{o}ssbauer sources with a much narrower line width, it is possible to
lower the minimally detectable value of anisotropy at least by three orders
of magnitude. Therefore, a repetition of the experiment [27] would now
be interesting.

\bigskip
\bigskip

\noindent
{\bf 2.3 Modification of fundamental relativistic
equations in the anisotropic space}

\bigskip

All fundamental relativistic equations are invariant under the 
transformations of the Poincar\'e group, the isometry group of
Minkowski space. If the event space is described by the Finslerian
metric (6), then the complete inhomogeneous group of its isometries
turns out to be an 8-parameter group. In this a case, the fundamental 
relativistic equations must be modified in accordance with the
requirement of invariance under this group. 

The requirement just formulated represents a generalization of the special
principle of relativity for the locally anisotropic space-time. Since the
8-parameter group in question is incorporated in the conformal group 
and the standard relativistic equations relating to massless
particles are conformally invariant, then only such equations continue to work
in the locally anisotropic space as well. The remaining relativistic equations
, in particular, the standard equations of relativistic mechanics are neither
conformally invariant nor invariant under the transformations
belonging to the 8-parameter linear subgroup of the conformal
group. Thus, such equations need modification. We now consider the
required modification [30] for the equations of mechanics. 

The equations of relativistic mechanics, which satisfy the special principle of
relativity for the locally anisotropic space, can be obtained if in the action
integral 
\be \lb{17}
S=-mc\int\limits_a^b\,ds
\ee
we replace the Minkowskian expression for $ds$ by the Finslerian expression
(6). As a result, the Lagrangian function corresponding to a free particle
in the locally anisotropic space, takes the form 
\be \lb{18}
L = -mc^2\left (\frac{1-\mbf v\nub /c}{\sqrt{1-\mbf v^{\,2}/c^2}}\right )^r
\,\sqrt{1-\mbf v^{\,2}/c^2}\,.
\ee
This Lagrangian leads to the following expressions for the momentum 
$\mbf p=\partial L/\partial\mbf v$ and the energy $E=\mbf p\mbf v-L$ 
of a relativistic particle 
\be \lb{19}
E = \frac{mc^2}{\sqrt{1-\mbf v^{\,2}/c^2}}\left (\frac{1-\mbf v\nub /c}{
\sqrt{1-\mbf v^{\,2}/c^2}}\right )^r\,\left [1-r+r\,\frac{1-\mbf v^{\,2}/c^2}{
1-\mbf v\nub /c}\right ]\,,
\ee
\be \lb{20}
\mbf p = \frac{mc}{\sqrt{1-\mbf v^{\,2}/c^2}}\left (\frac{1-\mbf v\nub /c}{
\sqrt{1-\mbf v^{\,2}/c^2}}\right )^r\,\left [(1-r)\mbf v/c+
r\nub\,\frac{1-\mbf v^{\,2}/c^2}{
1-\mbf v\nub /c}\right ]\,.
\ee
It can be verified by direct substitution that energy and momentum are
related by the relation
\be \lb{21} 
{\left [\frac{(E/c-\mbf p\nub )^2}{E^2/c^2- {\mbf p}^{\,2}}\right ]}^{-r}\,
(E^2/c^2-\mbf p^{\,2})=m^2c^2(1-r)^{(1-r)}(1+r)^{(1+r)}\,.
\ee
This relation determines the square of the Finslerian length of the
4-momentum $p\,.$ In passing from one inertial frame to another its
components $p^0=E/c$ and $\mbf p$ must transform such as to guarantee
invariance of the form (21). We have shown above that the invariance
of the Finslerian metric (6) is established by the generalized Lorentz
transformations (7). From the comparison of (21) and (6), the
invariance of (21) results from the transformations 
\be \lb{22}
p'^i=D^{-1}\,R_j^i\,L_k^j\,p^k\,,
\ee
where the matrices $L_k^j$ and $R_j^i$ are the same as in (7), while 
\be \lb{23}
D^{-1}=\left(\frac{1-\mbf v\nub /c}{\sqrt{1-{\mbf v}^{\,2}/c^2}}
\right )^{-r}I\,.            
\ee
Thus, under generalized Lorentz transformations the scale
transformation (23) for momenta is inverse to the corresponding scale
transformation (9) for the coordinates of events. Consequently, the
phase of a plane wave is an invariant of the generalized Lorentz
transformations. 

Eq.\ (19) determines the dependence of the energy $E$ of a free particle,
present in the anisotropic space, on both the magnitude and the direction of its
velocity $\mbf v\,.$ At $\mbf v=0$ the energy reaches its absolute minimum 
$E_0=mc^2\,.$  As
regards the momentum $\mbf p\,,$ its direction, according to (20),  does not
coincide with the direction of the velocity of a massive particle. Even
in the case $\mbf v=0\,,$ the momentum of a particle does not vanish; there
remains a ``rest momentum'' $\mbf p_0=rmc\nub \,$ . Massless particles have no
such property; for them, as in SR, $v=c$ and $E^2/c^2-{\mbf p}^{\,2}=0\,.$ 

In the space of 4-momenta $p$ the relation (21) is the equation of mass shell.
It appears as a deformed two-sheeted hyperboloid inscribed into a cone 
$p^{0\,2}-{\mbf p}^{\,2}=0\,.$
For the upper sheet of such a ``hyperboloid'' $p^0$ reaches its
absolute minimum $p^0_{min}=E_0/c=mc$ at $\mbf p=\mbf p_0=rmc\nub \,.$
For the lower sheet, $p^0$ reaches its absolute maximum
$p^0_{max}=-mc$ at $\mbf p=-rmc\nub \,.$
In order to display the mass shell graphically, let us introduce the
dynamic 4-velocity $u=p/mc$ in place of $p\,.$ We also put $c=1$ and choose
the coordinate axes such that $\nub =(1,0,0)\,.$ If we confine our
consideration to the case of two-dimensional motion and use polar
coordinates: $0\le v<1\,; \,\,0\le\alpha <2\pi\,,$ in which $\mbf
v=(v\cos\alpha\,,
v\sin\alpha\,, 0)\,,$
then, according to (21), (20), (19), the equation of (the positive
frequency part of) the mass shell 
\be \lb{24}
{\left[\frac{(u^0-u^1)^2}{(u^0)^2-(u^1)^2-(u^2)^2}\right]}^{-r}\,
\left[{(u^0)^2-(u^1)^2-(u^2)^2}\right]=(1-r)^{(1-r)}(1+r)^{(1+r)}
\ee
can be written in the following parametric  form (parameters $v\,, \alpha $)
\be \lb{25}
\left.
\begin{array}{ll}
u^1=\left.{\left(\frac{1-v\cos\alpha}{\sqrt{1-v^2}}\right)}^r\left[(1-r)v\cos\alpha
+\frac{r(1-v^2)}{1-v\cos\alpha}\right]\right/\sqrt{1-v^2}&,\\
                                                         & \\
u^2={\left(\frac{1-v\cos\alpha}{\sqrt{1-v^2}}\right)}^r\frac{(1-r)v\sin\alpha}{\sqrt{1-v^2}}\
\ ,& \\
   & \\
u^0=\left.{\left(\frac{1-v\cos\alpha}{\sqrt{1-v^2}}\right)}^r\left[1-r
+\frac{r(1-v^2)}{1-v\cos\alpha}\right]\right/\sqrt{1-v^2}\ \ . &
\end{array}
\right.
\ee
The results of calculations using (25) are presented in Fig.\,2\,. As,
according to (24), the mass shell is symmetric with respect
to the plane $u_2=0\,,$ the range of variation of the angle 
$\alpha $ was limited in Fig. 2 by the condition $0\le\alpha\le\pi\,.$ 

Being an intrinsic property of space, anisotropy is independent of the
magnitude of relative velocities. Therefore, also nonrelativistic
mechanics as a whole is different from the Newtonian case. In fact, in
the nonrelativistic limit the following expressions are obtained from
(19) and (20)
\be \lb{26} 
E=mc^2+(1-r)\frac{m{\mbf v}^{\,2}}{2}+r(1-r)\frac{m(\mbf v\nub )^2}{2}\,\,,
\ee
\be \lb{27}
\mbf p=rmc\nub +(1-r)m\mbf v+r(1-r)m(\mbf v\nub )\nub \,\,.
\ee
Since within the framework of nonrelativistic mechanics the rest mass $m$ is
an additive quantity, the occurrence of the constant terms $mc^2$ and $rmc\nub $
in (26) and (27) does not affect the conservation laws and the equations of
motion. As a result, these terms can be omitted, and the kinetic energy
and kinetic momentum, read off from (26) and (27), are 
\be \lb{28}
T={1\over 2}\,{\goth M}_{\alpha\beta}\,v^\alpha v^\beta \,,\qquad
p_\alpha ={\goth M}_{\alpha\beta}\,v^\beta \,,
\ee
where
\be \lb{29} 
{\goth M}_{\alpha\beta}=m(1-r)({\delta}_{\alpha\beta}+r{\nu}_\alpha\,{\nu}_
\beta )\,.  
\ee
Differentiating the second relation in (28) with respect to time, and using as
definition of force the derivative of momentum, we find that Newton's second
law in anisotropic space has the form 
\be \lb{30}
{\goth M}_{\alpha\beta}\,a^\beta =F_\alpha \,\, \ee \\ $(\alpha = 1, 2, 3)$.
Thus the inertial properties of a nonrelativistic particle in
anisotropic space is specified by a tensor of inertial mass (29);
its motion is analogous to the motion of a quasiparticle in a
crystalline medium. Equations as (30) were also discussed in the
framework of nonrelativistic Machian theories ([31]).

Irrespective of which closed physical system is considered  -- 
nonrelativistic or relativistic -- , according to Noether's first
theorem [32], there exist eight additive integrals of motion
corresponding to the eight independent symmetry transformations of the
space-time (6). Conservation of the total 4-momentum follows from the 
4-parameter translational symmetry of (6); conservation of three more 
quantities determining the coordinates of the center of inertia
corresponds to the 3-parameter generalized Lorentz symmetry of (6);
and, finally, conservation of the projection of the
total angular momentum of the system onto the preferred direction $\nub $
corresponds to the symmetry of (6) with respect to rotations about $\nub \,.$ 

The conservation law of {\em total momentum} manifests itself
differently in isotropic and anisotropic spaces. As an elementary
example, consider the elastic collision of two particles in isotropic
space, one of which at first was at rest. The conservation law of
total momentum then makes the tracks of the particle coplanar. For the
same process but now in anisotropic space, where the directions of
velocities and, hence, of the tracks of particles do not coincide
with the directions of their momenta, the conservation law of total momentum
does not lead to the fact that all the three tracks must necessarily lie in the
same plane. However, since the amount of the deviation from
coplanarity is a function of the magnitude of space anisotropy,
possible effects of noncoplanarity should be searched for in regions
where the magnitude of local anisotropy is significantly greater than
its mean value ( i.e.\ greater than $10^{-10}$ ). Such
a situation may obtain in the vicinity of very large masses, for
example, near the Sun. It seems reasonable to test this assumption with a
corresponding detector on a space vehicle able to identify elementary
events with nonstandard kinematics. 

Having mentioned that the magnitude of local anisotropy near massive bodies is
greater than that farther away, we thus strayed from the model
of space whose anisotropy is constant - always and everywhere. Introducing a 
{\em field} of anisotropy against the background of space-time curved
by matter would be a further step in the construction of a theory of
locally anisotropic space-time. We will turn to this problem in
Section 6. In the next Section, we consider another type of a
possible breaking of local isotropy, viz.\ a complete loss of symmetry with
respect to the group of 3D rotations.\\ 

\newpage
\noindent
{\bf 3. FINSLERIAN METRIC OF A FLAT SPACE-TIME\\
\phantom{Aa}WITH ENTIRELY BROKEN 3D ISOTROPY}

\bigskip
\bigskip
 
It was demonstrated in Section 2 that the homogeneous isometry group of a flat
space-time with a partially broken local isotropy, i.e.\ the homogeneous
isometry group of the space (6), is a 4-parameter group and includes,
apart from 3-parameter generalized Lorentz transformations (7), the 
1-parameter group of rotations about the preferred direction $\nub
\,.$ We now try to construct a geometric model of a flat space-time, the
homogeneous isometry group of which would only consist of noncompact 
3-parameter transformations of relativistic symmetry. For the solution of
this problem, the following observation is important. 

Consider the metric (6) in the limiting case $r=1\,.$ In this case 
\be \lb {31}
ds=dx_0-\nub d\mbf x\,,
\ee
and since $\nub = const\,,$ $ds$ turns out to be a total differential;
the action (17) for a free particle of mass $m$ is no longer dependent on the
shape of the world line connecting the points $a$ and $b\,.$ All this 
means that at $r=1$ a massive particle loses its inertia. This can be
illustrated by Eq.\ (29) which determines the inertial mass tensor 
${\,\goth M}_{\alpha \beta}\,,$ and also by Eqs.\ (19) and (20) which 
determine the dependence of the energy $E$ and the momentum $\mbf p$
on the particle velocity $\mbf v\,.$ From these formulae,
at $r=1\,,$ it follows that $\,{\goth M}_{\alpha \beta}=0\,$ while $E$ and $\mbf
p$ 
become no longer
dependent on $\mbf v$ and become equal to the corresponding constants $mc^2$ and
$mc\nub \,.$ At $r=1\,,$ apart from inertness, the notion of spatial extension
disappears, which is due to the absence of a light cone and, hence, of the
possibility itself for determining spatial distances with the aid of exchange
of light signals. As a result, in the space-time (31) there remains a single
physical characteristic --- time duration $ds$ and it should be regarded as an
interval of absolute time. 

Since the ``metric'' (31) is a special case of the metric (6), all 
transformations leaving invariant the metric (6) leave invariant the
``metric'' (31) as well. This likewise applies to the transformations
(7), if in them $r=1$ is set. It can
readily be seen, however, that in comparison with (6) the ``metric'' (31)
possesses an additional symmetry. Its existence becomes evident after
substitution of the variables 
$$\nu _1\,x_1 \to x_1\,,\quad \nu _2\,x_2 \to x_2\,,\quad \nu _3\,x_3 \to
x_3\,;\qquad \nu _1\,, \nu _2\,, \nu _3\,\ne 0\,,
$$
as a result of which Eq.\ (31) takes the form 
\be \lb{32}
ds=dx_0-dx_1-dx_2-dx_3\,.
\ee
Thus our observation (suggested by invariance of the expression (4) under (2)
at $r=1$) consists in the fact that the additional symmetry of the 1-form (32)
is realized as its invariance under the following three independent
1-parameter groups of transformations 
\be \lb{33}
\left\{
\begin{array}{lllcll}
x'_0=e^{-\alpha _1}&(&x_0\,\cosh \alpha _1&-&x_1\,\sinh \alpha _1&)\\
x'_1=e^{-\alpha _1}&(\,-&x_0\,\sinh \alpha _1&+&x_1\,\cosh \alpha _1&)\\
x'_2=e^{-\alpha _1}&(&x_2\,\cosh \alpha _1&+&x_3\,\sinh \alpha _1&)\\
x'_3=e^{-\alpha _1}&(&x_2\,\sinh \alpha _1&+&x_3\,\cosh \alpha _1&)\,,\\
\end{array}
\right.
\ee
\be \lb{34}
\left\{
\begin{array}{lllcll}
x'_0=e^{-\alpha _2}&(&x_0\,\cosh \alpha _2&-&x_2\,\sinh \alpha _2&)\\
x'_1=e^{-\alpha _2}&(&x_1\,\cosh \alpha _2&+&x_3\,\sinh \alpha _2&)\\
x'_2=e^{-\alpha _2}&(\,-&x_0\,\sinh \alpha _2&+&x_2\,\cosh \alpha _2&)\\
x'_3=e^{-\alpha _2}&(&x_1\,\sinh \alpha _2&+&x_3\,\cosh \alpha _2&)\,,\\
\end{array}
\right.
\ee
\be \lb{35}
\left\{
\begin{array}{lllcll}
x'_0=e^{-\alpha _3}&(&x_0\,\cosh \alpha _3&-&x_3\,\sinh \alpha _3&)\\
x'_1=e^{-\alpha _3}&(&x_1\,\cosh \alpha _3&+&x_2\,\sinh \alpha _3&)\\
x'_2=e^{-\alpha _3}&(&x_1\,\sinh \alpha _3&+&x_2\,\cosh \alpha _3&)\\
x'_3=e^{-\alpha _3}&(\,-&x_0\,\sinh \alpha _3&+&x_3\,\cosh \alpha _3&)\,.\\
\end{array}
\right.
\ee
Note that each of the groups (33)--(35) is represented by nonorthogonal
transformations. For example, the transformations (33) change the angle
between the axes $x_2$ and $x_3\,.$  Apart from the 1-form (32), the 
transformations (33), (34) and (35), respectively leave invariant the 1-forms 
\be \lb{36}
ds=dx_0-dx_1+dx_2+dx_3\,;
\ee

\be \lb{37}
ds=dx_0+dx_1-dx_2+dx_3
\ee
and
\be \lb{38}
ds=dx_0+dx_1+dx_2-dx_3\,.
\ee

Let us make now the following step by introducing into (33)--(35) a
dependence of the dilatations $e^{-\alpha _1}\,, e^{-\alpha _2}\,,$
and $e^{-\alpha _3}$ on three independent
parameters $r_1\,, r_2 $ and $r_3\,,$ respectively. As a result of such
generalization we arrive at three independent 1-parameter groups of
transformations, corresponding to eqs. (33)--(35) but in which the
factors $e^{-\alpha_i}$ are replaced by $e^{-r_i\cdot\alpha_i}$,
respectively ($\alpha _i$ still act as  group parameters). 
These transformations act on the 1-forms (32), (36--38) in the
following way
\be \lb{39}
\left\{
\begin{array}{lcll}
dx'_0-dx'_1-dx'_2-dx'_3&=&e^{(1-r_1)\,\alpha _1}&(dx_0-dx_1-dx_2-dx_3)\\
dx'_0-dx'_1+dx'_2+dx'_3&=&e^{(1-r_1)\,\alpha _1}&(dx_0-dx_1+dx_2+dx_3)\\
dx'_0+dx'_1-dx'_2+dx'_3&=&e^{-(1+r_1)\,\alpha _1}&(dx_0+dx_1-dx_2+dx_3)\\
dx'_0+dx'_1+dx'_2-dx'_3&=&e^{-(1+r_1)\,\alpha
_1}&(dx_0+dx_1+dx_2-dx_3)\,,\\
\end{array}
\right.
\ee
\be \lb{40}
\left\{
\begin{array}{lcll}
dx'_0-dx'_1-dx'_2-dx'_3&=&e^{(1-r_2)\,\alpha _2}&(dx_0-dx_1-dx_2-dx_3)\\
dx'_0-dx'_1+dx'_2+dx'_3&=&e^{-(1+r_2)\,\alpha _2}&(dx_0-dx_1+dx_2+dx_3)\\
dx'_0+dx'_1-dx'_2+dx'_3&=&e^{(1-r_2)\,\alpha _2}&(dx_0+dx_1-dx_2+dx_3)\\
dx'_0+dx'_1+dx'_2-dx'_3&=&e^{-(1+r_2)\,\alpha
_2}&(dx_0+dx_1+dx_2-dx_3)\,,\\
\end{array}
\right.
\ee
\be \lb{41}
\left\{
\begin{array}{lcll}
dx'_0-dx'_1-dx'_2-dx'_3&=&e^{(1-r_3)\,\alpha _3}&(dx_0-dx_1-dx_2-dx_3)\\
dx'_0-dx'_1+dx'_2+dx'_3&=&e^{-(1+r_3)\,\alpha _3}&(dx_0-dx_1+dx_2+dx_3)\\
dx'_0+dx'_1-dx'_2+dx'_3&=&e^{-(1+r_3)\,\alpha _3}&(dx_0+dx_1-dx_2+dx_3)\\
dx'_0+dx'_1+dx'_2-dx'_3&=&e^{(1-r_3)\,\alpha
_3}&(dx_0+dx_1+dx_2-dx_3)\,,\\
\end{array}
\right.
\ee
Since, according to (39--41), there occur only scale transformations
of the four introduced 1-forms,
we try to seek the metric for the flat Finslerian space-time (with an entirely
broken symmetry with respect to 3D rotations) in the form 
\be \lb {42}
\begin{array}{rl}
ds=&(dx_0-dx_1-dx_2-dx_3)^a(dx_0-dx_1+dx_2+dx_3)^b\\
\times &(dx_0+dx_1-dx_2+dx_3)^c(dx_0+dx_1+dx_2-dx_3)^d\,,\\
\end{array}
\ee
where $a\,,\ b\,,\ c\,,\ d\ $ are some constants for the determination
of which the following four conditions must be fulfilled: (i) the
metric (42) should be a
homogeneous function of the coordinate differentials of the first degree of
homogeneity; and (ii)--(iv) the metric (42) should remain invariant under the
transformations belonging to any of the three independent groups (39--41). 
These conditions lead to a system of four equations 
$$ 
\left\{
\begin{array}{rccrccrccrccl}
&a&+&&b&+&&c&+&&d&=&1\\
(1-r_1)&a&+&(1-r_1)&b&-&(1+r_1)&c&-&(1+r_1)&d&=&0\\
(1-r_2)&a&-&(1+r_2)&b&+&(1-r_2)&c&-&(1+r_2)&d&=&0\\
(1-r_3)&a&-&(1+r_3)&b&-&(1+r_3)&c&+&(1-r_3)&d&=&0\,.\\
\end{array}
\right.
$$
The determinant of the given system is equal to $-16$ while its solution is of
the form
$$
\begin{array}{ll}
a=(1+r_1+r_2+r_3)\,/\,4\,,&b=(1+r_1-r_2-r_3)\,/\,4\,,\\
c=(1-r_1+r_2-r_3)\,/\,4\,,&d=(1-r_1-r_2+r_3)\,/\,4\,.\\
\end{array}
$$
Thus, taking into account (42), we obtain the required expression [19] for the
metric of the flat locally anisotropic space-time with entirely broken
rotational symmetry.\footnote{The
  general form of this line element is $ds = \{(a_i dx^i)^{1  +
    \alpha} (b_j dx^j)^{1 + \beta} (c_k dx^k)^{1 + \gamma} (d_l
  dx^l)^{1 + \delta})\}^{1/4}$ with $ ~ \alpha + \beta + \gamma +\delta = 0
  $. This is an example for the ``4th square root of a differential 
expression of fourth degree'' announced by Riemann as the second simplest line
  element of what later became known as Finsler spaces [20].} 
\be \lb{43}
\begin{array}{rl}
ds=&(dx_0-dx_1-dx_2-dx_3)^{(1+r_1+r_2+r_3)\,/\,4}\\
\times &(dx_0-dx_1+dx_2+dx_3)^{(1+r_1-r_2-r_3)\,/\,4}\\
\times &(dx_0+dx_1-dx_2+dx_3)^{(1-r_1+r_2-r_3)\,/\,4}\\
\times &(dx_0+dx_1+dx_2-dx_3)^{(1-r_1-r_2+r_3)\,/\,4}\,.\\
\end{array}
\ee
The anisotropy of the Finslerian space (43) is now specified by even the three
parameters $r_1\,,\ r_2\,,\ r_3$ which satisfy the conditions 
\be \lb{44}
\begin{array}{ll}
1+r_1+r_2+r_3>0\,,&1+r_1-r_2-r_3>0\,,\\
1-r_1+r_2-r_3>0\,,&1-r_1-r_2+r_3>0\,.\\
\end{array}
\ee
These conditions ensure the fact that the section of a light cone by
hyperplane $dx_0=const$ is a closed convex surface. This, in turn, ensures the
applicability of the procedure of exchange of light signals for determining
3D distances. 

According to (44), the permissible values of the parameters $r_1\,,\ r_2\,,\
r_3$
fill the inner region of a regular tetrahedron with the vertices at the points 
$$
\begin{array}{ll}
(r_1=1\,,r_2=1\,,r_3=1)\,;&(r_1=1\,,r_2=-1\,,r_3=-1)\,;\\
(r_1=-1\,,r_2=1\,,r_3=-1)\,;&(r_1=-1\,,r_2=-1\,,r_3=1)\,.\\
\end{array}
$$
At these four points the metric (43) degenerates into the
corresponding 1-forms (32), (36--38), i.e.\ into the total
differentials of absolute time. We now recall that the metric (6) of
the flat locally anisotropic space-time with the
partially broken 3D isotropy also degenerates, at $r=1\,,$ into the total
differential of absolute time. This suggests that absolute time  is not a
stable degenerate state of space-time and (as a result of the geometric phase
transition) may turn either into the partially anisotropic space-time (6) or
into the entirely anisotropic space-time (43). Such a phase transition
is could be interpreted as an ``act of creation'' of a 3D space. In the
passage to (6) there occurs a 3D space with locally Euclidean geometry while
in the passage to (43) there occurs, as will be shown below, a flat 3D space
with non-Euclidean geometry. Thus, absolute time plays the role of a
connecting link by which a principle of correspondence is satisfied for the
Finslerian spaces (6) and (43). 

In order to better understand the role of the parameters
$r_1\,,r_2\,,r_3\,,$ we put in the metric (43) $\,\,dx_2=dx_3=0\,.$ As
a result, it turns out that $$ds=[(dx_0-dx_1)^2\,/\,(dx_0^2-dx_1^2)]^
{r_1\,/\,2}\,\sqrt {dx_0^2-dx_1^2}\,\,.$$ 
In (43), we now put $\,dx_1=dx_3=0\,.$ Then we obtain $ds$ in the form 
$$
ds=[(dx_0-dx_2)^2\,/\,(dx_0^2-dx_2^2)]^{r_2\,/\,2}\,\sqrt {dx_0^2-dx_2^2}\,\,.
$$ 
Similarly, by putting in (43) $\,\,dx_1=dx_2=0\,,$ we arrive at the metric 
$$
ds=[(dx_0-dx_3)^2\,/\,(dx_0^2-dx_3^2)]^{r_3\,/\,2}\,\sqrt {dx_0^2-dx_3^2}\,\,.
$$ 
Each of these three expressions is idential with the expression (4)
which represents the metric of a 2D anisotropic space-time. Therefore, in a
sense, the parameters $\,r_1\,,r_2\,,r_3\,$ characterize the
anisotropy along the corresponding axes $\,x_1\,,x_2\,,x_3\,.$
However, space-time (43) is such that it remains anisotropic even at 
$r_1=r_2=r_3=0\,.$ 

In summing up, we see that the 2D anisotropic metric (4) admits two 
independent ways of generalization to four dimensions. The first way
leads to the partially anisotropic Finslerian 4D metric (6) and the
second one to the totally anisotropic Finslerian 4D metric (43). 

\bigskip
\bigskip

\noindent
{\bf 4. HOMOGENEOUS GROUP OF RELATIVISTIC SYMMETRY\\
\phantom{Aa}OF THE ENTIRELY ANISOTROPIC SPACE-TIME}

\bigskip
\bigskip
 
Consider an homogeneous isometry group of the flat space-time (43). By its
construction the metric (43) is an invariant of the three independent
1-parameter group of the transformations (39--41). In their infinitesimal
form, the transformations belonging to these groups appear as 
$$
\left\{
\begin{array}{lcl}
dx_0&=&(-r_1x_0-x_1)d\alpha _1\\
dx_1&=&(-r_1x_1-x_0)d\alpha _1\\
dx_2&=&(-r_1x_2+x_3)d\alpha _1\\
dx_3&=&(-r_1x_3+x_2)d\alpha _1\,,\\
\end{array}
\right.
\qquad
\left\{
\begin{array}{lcl}
dx_0&=&(-r_2x_0-x_2)d\alpha _2\\
dx_1&=&(-r_2x_1+x_3)d\alpha _2\\
dx_2&=&(-r_2x_2-x_0)d\alpha _2\\
dx_3&=&(-r_2x_3+x_1)d\alpha _2\,,\\
\end{array}
\right.
$$
$$
\left\{
\begin{array}{lcl}
dx_0&=&(-r_3x_0-x_3)d\alpha _3\\
dx_1&=&(-r_3x_1+x_2)d\alpha _3\\
dx_2&=&(-r_3x_2+x_1)d\alpha _3\\
dx_3&=&(-r_3x_3-x_0)d\alpha _3\,.\\
\end{array}
\right.
$$ 
It can easily be verified that the corresponding generators 
$$
X_1=-r_1x_i\,\partial \,/\,\partial x_i-(x_1\,\partial \,/\,\partial x_0+
x_0\,\partial \,/\,\partial x_1)+(x_2\,\partial \,/\,\partial x_3+
x_3\,\partial \,/\,\partial x_2)\,,
$$
$$
X_2=-r_2x_i\,\partial \,/\,\partial x_i-(x_2\,\partial \,/\,\partial x_0+
x_0\,\partial \,/\,\partial x_2)+(x_1\,\partial \,/\,\partial x_3+
x_3\,\partial \,/\,\partial x_1)\,,
$$
$$
X_3=-r_3x_i\,\partial \,/\,\partial x_i-(x_3\,\partial \,/\,\partial x_0+
x_0\,\partial \,/\,\partial x_3)+(x_1\,\partial \,/\,\partial x_2+
x_2\,\partial \,/\,\partial x_1)\,\,
$$ 
commute, i.e.\ $[X_\alpha X_\beta ]=0\,.$ It thus appears that the homogeneous
3-parameter noncompact isometry group, i.e. the relativistic symmetry
group of the space-time (43) is Abelian and any of its
elements can be obtained by multiplying (in an arbitrary order) the
transformations (39--41). Having made such multiplication we arrive
at the required 3-parameter transformations 
\be \lb{45}
x'_i=D\,L_{ik}\,x_k\,.
\ee
Here $\ D=\exp (-r_1\,\alpha _1-r_2\,\alpha _2-r_3\,\alpha _3\,)\,;$
\  the matrices 
$$
L_{ik}=\left (
\begin{array}{rrrr}
\cal A&-\cal B&-\cal C&-\cal D\\
-\cal B&\cal A&\cal D&\cal C\\
-\cal C&\cal D&\cal A&\cal B\\
-\cal D&\cal C&\cal B&\cal A\\
\end{array}
\right )
$$ 
are unimodular, whereby 
$$
{\cal A}=\cosh \alpha _1\cosh \alpha _2\cosh \alpha _3+
\sinh \alpha _1\sinh \alpha _2\sinh \alpha _3\,,
$$
$$
{\cal B}=\cosh \alpha _1\sinh \alpha _2\sinh \alpha _3+
\sinh \alpha _1\cosh \alpha _2\cosh \alpha _3\,,
$$
$$
{\cal C}=\cosh \alpha _1\sinh \alpha _2\cosh \alpha _3+
\sinh \alpha _1\cosh \alpha _2\sinh \alpha _3\,,
$$
$$
{\cal D}=\cosh \alpha _1\cosh \alpha _2\sinh \alpha _3+
\sinh \alpha _1\sinh \alpha _2\cosh \alpha _3\,;
$$
and $\,\alpha _1\,,\alpha _2\,,\alpha _3\,$ are the group
parameters. The transformations inverse to (45) can be obtained if we
make the substitution 
$$\alpha _1\ \to\ -\alpha _1\,,\qquad \alpha _2\ \to\ -\alpha _2\,,
\qquad \alpha _3\ \to\ -\alpha _3\,.
$$ 
As a result 
\be \lb{46}
x_i=D^{-1}\,L^{-1}_{ik}\,x'_k\,,
\ee
where
$$
L^{-1}_{ik}=\left (
\begin{array}{rrrr}
\tilde{\cal A}&-\tilde{\cal B}&-\tilde{\cal C}&-\tilde{\cal D}\\
-\tilde{\cal B}&\tilde{\cal A}&\tilde{\cal D}&\tilde{\cal C}\\
-\tilde{\cal C}&\tilde{\cal D}&\tilde{\cal A}&\tilde{\cal B}\\
-\tilde{\cal D}&\tilde{\cal C}&\tilde{\cal B}&\tilde{\cal A}\\
\end{array}
\right )\,,
$$
\be \lb{47}
\tilde{\cal A}=\cosh \alpha _1\cosh \alpha _2\cosh \alpha _3-
\sinh \alpha _1\sinh \alpha _2\sinh \alpha _3\,,
\ee
\be \lb{48}
\tilde{\cal B}=\cosh \alpha _1\sinh \alpha _2\sinh \alpha _3-
\sinh \alpha _1\cosh \alpha _2\cosh \alpha _3\,,
\ee
\be \lb{49}
\tilde{\cal C}=\sinh \alpha _1\cosh \alpha _2\sinh \alpha _3-
\cosh \alpha _1\sinh \alpha _2\cosh \alpha _3\,,
\ee
\be \lb{50}
\tilde{\cal D}=\sinh \alpha _1\sinh \alpha _2\cosh \alpha _3-
\cosh \alpha _1\cosh \alpha _2\sinh \alpha _3\,.
\ee

Since the relativistic symmetry transformations (45) have the same meaning as
the Lorentz transformations, it is helpful to use as group parameters,
in place of $\ \alpha _1\,,\ \alpha _2\,,\ \alpha _3\,,\ $ the components
$\,v_1\,,\ v_2\,,\ v_3\,$ of the velocity of the primed frame. In
order to obtain the necessary relations it is sufficient to put
$\,x'_1=x'_2=x'_3=0\,$ in (46). As a result 
\be \lb{51}
v_1=\frac{x_1}{x_0}=-\frac{\tilde{\cal B}}{\tilde{\cal A}}\,,\
v_2=\frac{x_2}{x_0}=-\frac{\tilde{\cal C}}{\tilde{\cal A}}\,,\
v_3=\frac{x_3}{x_0}=-\frac{\tilde{\cal D}}{\tilde{\cal A}}\,.
\ee
Taking into account (47--50), we can rewrite these formulae as follows 
$$v_1=(\tanh\alpha _1-\tanh\alpha_2\tanh\alpha _3)/(
1-\tanh\alpha_1\tanh\alpha_2\tanh\alpha_3)\,,
$$
$$v_2=(\tanh\alpha _2-\tanh\alpha_1\tanh\alpha _3)/(
1-\tanh\alpha_1\tanh\alpha_2\tanh\alpha_3)\,,
$$
$$v_3=(\tanh\alpha _3-\tanh\alpha_1\tanh\alpha _2)/(
1-\tanh\alpha_1\tanh\alpha_2\tanh\alpha_3)\,.
$$ 

Now find the inverse relations, i.e.\ express $\,\alpha_1\,,\ \alpha_2\,,\
 \alpha_3\,$ in terms of $\,v_1\,,\ v_2\,,\ v_3\,.$ This is easy to do
 if the following formulae are used 
$$
1-v_1-v_2-v_3=\frac{(1-\tanh\alpha _1)(1-\tanh\alpha _2)(1-\tanh\alpha _3)}{
(1-\tanh\alpha _1\tanh\alpha _2\tanh\alpha _3)}\,,
$$
$$
1-v_1+v_2+v_3=\frac{(1-\tanh\alpha _1)(1+\tanh\alpha _2)(1+\tanh\alpha _3)}{
(1-\tanh\alpha _1\tanh\alpha _2\tanh\alpha _3)}\,,
$$
$$
1+v_1-v_2+v_3=\frac{(1+\tanh\alpha _1)(1-\tanh\alpha _2)(1+\tanh\alpha _3)}{
(1-\tanh\alpha _1\tanh\alpha _2\tanh\alpha _3)}\,,
$$
$$
1+v_1+v_2-v_3=\frac{(1+\tanh\alpha _1)(1+\tanh\alpha _2)(1-\tanh\alpha _3)}{
(1-\tanh\alpha _1\tanh\alpha _2\tanh\alpha _3)}\,.
$$ 
As a result we obtain 
$$
\alpha _1\,=\frac{1}{4}\ln \frac{(1+v_1-v_2+v_3)(1+v_1+v_2-v_3)}
{(1-v_1-v_2-v_3)(1-v_1+v_2+v_3)}\,,
$$
$$
\alpha _2\,=\frac{1}{4}\ln \frac{(1-v_1+v_2+v_3)(1+v_1+v_2-v_3)}
{(1-v_1-v_2-v_3)(1+v_1-v_2+v_3)}\,,
$$
$$
\alpha _3\,=\frac{1}{4}\ln \frac{(1-v_1+v_2+v_3)(1+v_1-v_2+v_3)}
{(1-v_1-v_2-v_3)(1+v_1+v_2-v_3)}\,.
$$ 

Since $\,v_1\,,\ v_2\,,\ v_3\,$ by definition are components of the coordinate
velocity of the primed frame and the light cone equation for the
entirely anisotropic event space (43) differs from the light cone equation of
Minkowski space, it is clear that in the entirely anisotropic space
an observable such as the magnitude of velocity no longer is
determinted by the Euclidean expression
$\,v=\sqrt{v_1^2+v_2^2+v_3^2}\,.$ In order to obtain the correct
formula for $\,v\,$ it is first necessary to formulate a procedure for 
synchronizing coordinate clocks, i.e.\ for determining the difference 
$ \Delta x_0 $ of the readings of coordinate clocks,
which correspond to simultaneous events at neighbouring points of the space
(43), and also to determine the observable distance between these points. 

\bigskip
\bigskip

\noindent
{\bf 5. \,\,3D GEOMETRY AND CLOCK SYNCHRONIZATION\\
\phantom{Aa}\,\,\,IN THE ENTIRELY ANISOTROPIC SPACE-TIME} 

\bigskip
\bigskip

According to the definition of the totally anisotropic metric (43), the range
of permissible values of  $\,dx_i\,$  is limited by the conditions 
\be \lb{52}
\left\{
\begin{array}{rcl}
dx_0-dx_1-dx_2-dx_3&\ge&0\\
dx_0-dx_1+dx_2+dx_3&\ge&0\\
dx_0+dx_1-dx_2+dx_3&\ge&0\\
dx_0+dx_1+dx_2-dx_3&\ge&0\,.\\
\end{array}
\right.
\ee
Being invariant under the relativistic transformations (45), these conditions
determine either a timelike interval between two events or an interval equal to
zero. The latter case corresponds to events related by a
light signal. Owing to the Abelian structure of the group (45) the cited
invariance of conditions (52) follows from the relations (39--41). Apart
from this, the transformations (45) leave invariant the sign of $\,dx_0\,.$ 

Now, let $\,dx_0>0\,.$ Then, in terms of the components $\,v_\alpha 
=dx_\alpha\,/dx_0\,$ of
the coordinate velocity, the conditions (52) can be rewritten as 
\be \lb{53}
\left\{
\begin{array}{rcl}
1-v_1-v_2-v_3&\ge&0\\
1-v_1+v_2+v_3&\ge&0\\
1+v_1-v_2+v_3&\ge&0\\
1+v_1+v_2-v_3&\ge&0\,.\\
\end{array}
\right.
\ee
The range of $\,v_\alpha \,$-values, limited by the conditions (53),
is represented in Fig.\,3\,. It forms a regular tetrahedron with its center
at the origin $\,o\,$ of a rectangular system of coordinates 
$\,v_1\,,v_2\,,v_3\,.$ The velocities corresponding to the timelike
intervals $\,ds\,$ fill the inner region of the tetrahedron while the 
velocities describing the propagation of light signals and guaranteing
$\,ds=0\,$ fill the surface of the tetrahedron. In comparison, we note
that in the case of Minkowski space, in place of (53), the
relativistically invariant constraint $\,1-\mbf v^{\,2}\ge 0$
obtains; i.e.\,, in place of the tetrahedron, a sphere of unit radius as
the range of permissible $\,v_\alpha \,$ values occurs.

Each face of the tetrahedron is described by one of the four equations 
\be \lb{54}
1-v_1-v_2-v_3=0\,;
\ee
\be \lb{55}
1-v_1+v_2+v_3=0\,;
\ee
\be \lb{56}
1+v_1-v_2+v_3=0\,;
\ee
\be \lb{57}
1+v_1+v_2-v_3=0\,,
\ee
and each of its six edges by a system of two equations chosen properly from
(54--57). The face $\,\Delta \Xi \Lambda \,$ is described by Eq.\ (54), 
the face $\,\Psi \Xi \Lambda \,,$ by (55), the face $\,\Psi \Delta \Xi
\,,$ by (56), the face $\,\Psi\Lambda \Delta \,,$ by (57), while for
example the edge $\,\Delta\Xi\,,$ by the  system of equations by (54) and (56)
etc. 

On the surface of the tetrahedron we mark 14 characteristic points: 
$\alpha \,, \beta \,, \gamma \,, \delta \,, \epsilon \,, \zeta \,,
\Gamma \,,$\\ $\Delta \,, \Theta \,, \Lambda \,, \Xi \,, \Phi \,, \Psi \,, 
\Omega \,.$ Let us represent the coordinates of these points in
the form of the rectangular components of the corresponding radius
vectors. In particular: $\,\overleftarrow{\alpha
o}=(1\,,0\,,0)\,;
\ \overleftarrow{\beta o}=(0\,,1\,,0)\,; \ \overleftarrow{\gamma
o}=(0\,,0\,,1)\,; \ \overleftarrow{\delta o}=(-1\,,0\,,0)\,;\ 
\overleftarrow{\epsilon o}=(0\,,-1\,,0)\,;\ 
\overleftarrow{\zeta o}=(0\,,0\,,-1)\,;\ 
\,\overleftarrow{\Delta o}=(-1\,,1\,,1)\,;\ 
\,\overleftarrow{\Lambda o}=(1\,,-1\,,1)\,;\
\,\overleftarrow{\Xi o}=(1\,,1\,,-1)\,;$ and 
$\,\overleftarrow{\Psi o}=(-1\,,-1\,,-1)\,.$ Similarly
the radius vector $\,\overleftarrow{\Gamma o}=(1/3\,,1/3\,,1/3)\,$ represents
the point 
$\,\Gamma \,.$ This point
is located at the center of the face $\,\Delta \Xi \Lambda \,$ and coincides
with the 
projection of
the vertex $\,\Psi \,$ onto this face. The radius vector
$\,\overleftarrow{\Omega
o}=(1/3\,,-1/3\,,-1/3)\,$
represents the point $\,\Omega \,$ located at the center of the face $\,\Psi \Xi
\Lambda \,$
and coinciding with the projection of the vertex $\,\Delta \,$ onto
it. Likewise, the radius vector
$\,\overleftarrow{\Phi o}=
(-1/3\,,1/3\,,-1/3)\,$ represents the point $\,\Phi \,$ located at the
center of the face $\,\Psi \Delta \Xi \,$ and coinciding with the projection 
of the vertex $\,\Lambda \,$ onto
this face. Finally, the radius vector $\,\overleftarrow{\Theta
o}=(-1/3\,,-1/3\,,1/3)\,$ 
represents the
point $\,\Theta \,$ located at the center of the face $\,\Psi \Lambda \Delta \,$
 and coinciding with the
projection of the vertex $\,\Xi \,$ onto this face. 

By these characteristic points on the tetrahedron surface this surface
is divided into twelve equal tetragons which, in turn,
are grouped into six pairs of mutually conjugate\footnote{with respect
  to a reflection operation at the origin} tetragons. Denoting the
reflection operation by a symbol $\ \longleftrightarrow \ \ ,$ we
obtain the following pairs$$
\begin{array}{lcl}
\Gamma \gamma \Delta \beta &\ \longleftrightarrow \ &\Psi \zeta \Omega
\epsilon \\
\Gamma \beta \Xi \alpha &\ \longleftrightarrow \ &\Psi \epsilon
\Theta \delta \\
\Gamma \alpha \Lambda \gamma &\ \longleftrightarrow \ &\Psi \delta
\Phi \zeta \\
\Omega \epsilon \Lambda \alpha &\ \longleftrightarrow \ &\Delta
\beta \Phi \delta \\
\Omega \alpha \Xi \zeta &\ \longleftrightarrow \ &\Delta \delta
\Theta \gamma \\
\Theta \gamma \Lambda \epsilon &\ \longleftrightarrow \ &\Xi
\zeta \Phi \beta \,.\\
\end{array}
$$ 
In accordance with the division of the tetrahedron surface, the full solid
angle $4\pi $ is also divided into six pairs of mutually conjugate
sectors. Each
of these sectors constitutes a tetrahedral solid angle which rests on the
corresponding tetragon and has its vertex at the origin of the coordinates.
Consider, for example, the sector $\,\Gamma \gamma \Delta \beta o\,$ which 
rests on the tetragon $\,\Gamma \gamma \Delta \beta \,.$ This
tetragon belongs to the face $\,\Delta \Xi \Lambda \,.$ Therefore, the
coordinates $\,v_1\,,v_2\,,v_3\,$
of any inner point of such a tetragon or of a point belonging to its boundary
satisfy Eq.\ (54), in which case the radius
vector $\,\overleftarrow{vo}=(\,v_1\,,v_2\,,v_3\,)\,$ represents the 
coordinate velocity of an initial
light ray propagating within the sector $\,\Gamma \gamma \Delta \beta
o\,$ since $\,ds=0\,$ (in virtue of (54)). Using Fig.\,3, and taking
into consideration the equation 
\be \lb{58}
1-{\tilde v}_1+{\tilde v}_2+{\tilde v}_3=0\,,
\ee
which describes the face $\,\Psi \Xi \Lambda \,,$ (\,cf.\,(55)\,) it
is easy to verify that the radius vector $\,\tilde{\overleftarrow{vo}}=
(\,{\tilde v}_1\,,{\tilde v}_2\,,{\tilde v}_3\,)\,$ with components 
\be \lb{59}
{\tilde v}_1=-\frac{v_1}{v_2+v_3-v_1}\,;\ {\tilde v}_2=
-\frac{v_2}{v_2+v_3-v_1}\,;\ {\tilde v}_3=-\frac{v_3}{v_2+v_3-v_1}\,,
\ee
where, according to (54), 
\be \lb{60}
v_1+v_2+v_3=1\,,
\ee
represents the coordinate velocity of a reflected light ray. Compared with the
initial ray such a ray has the opposite direction and propagates within the
sector $\,\Psi \zeta \Omega \epsilon o\,$ which rests on the tetragon $\,\Psi
\zeta \Omega
\epsilon \,.$ Formulae (59), (60) give a
one-to-one mapping of the tetragon $\,\Gamma \gamma \Delta \beta \,$ onto the
tetragon 
$\,\Psi \zeta \Omega \epsilon \,.$ The formulae
inverse to (59), (60) appear as 
\be \lb{61}
v_1=\frac{{\tilde v}_1}{{\tilde v}_1+{\tilde v}_2+{\tilde v}_3}\,;
\ v_2=\frac{{\tilde v}_2}{{\tilde v}_1+{\tilde v}_2+{\tilde v}_3}\,;
\ v_3=\frac{{\tilde v}_3}{{\tilde v}_1+{\tilde v}_2+{\tilde v}_3}\,,
\ee
where, according to (58), 
\be \lb{62}
{\tilde v}_2+{\tilde v}_3-{\tilde v}_1=-1\,.
\ee
These formulae also give a one-to-one mapping of the tetragon 
$\,\Psi \zeta \Omega\epsilon \,$ onto the tetragon $\,\Gamma \gamma
\Delta \beta \,.$ It is precisely in connection with the mappings
(59), (60) and (61), (62) that the
tetragons $\,\Gamma \gamma \Delta \beta \,$ and $\,\Psi \zeta \Omega \epsilon
\,$ 
(as well as the corresponding sectors) were called mutually conjugate above. 

The formulae, which relate the components of the
coordinate velocities of initial and reflected light rays are modified in the
passage from one pair of mutually conjugate sectors to another. This involves a
corresponding change in the formulae for the observables, the change
being such that the observables remain continuous at the boundaries separating
neighbouring sectors. This is confirmed by the Table
given below, in which formulae are collected which determine the observables
for each of the twelve sectors. For illustrative purposes we reproduce
here only the formulae pertaining to the sector $\,\Gamma \gamma
\Delta \beta o\,.$ The meaning of all the symbols involved in the
Table of observables should now have become obvious. 

In order to determine how the difference of the coordinates of two events in
the event space (43) is correlated with the observables, we use the Einstein
procedure implying exchange of light signals between points of 3D space. 

Let an initial event $\,I\,,$  with coordinates
$\,(\,0\,,0\,,0\,,0\,)\,,$ be involved in  the emission of a light signal, 
and another event $\,R\,$ with coordinates 
$\,(\,dx_0^{(1)}\,,dx_1\,,dx_2\,,dx_3\,)\,$ be involved in the
reflection of this signal. In addition, let the
$\,dx_1\,,dx_2\,,dx_3\,$ be such that the initial signal
propagates within the sector $\,\Gamma \gamma \Delta \beta o\,.$ Represent 
the components of the coordinate velocity of the initial signal in the form 
\be \lb{63}
v_1=\frac{dx_1}{dx_0^{(1)}}\,;\quad v_2=\frac{dx_2}{dx_0^{(1)}}\,;\quad
v_3=\frac{dx_3}{dx_0^{(1)}}\,.
\ee
Finally, let 
\be \lb{64}
 (\,dx_0^{(1)}+dx_0^{(2)}\,,0\,,0\,,0\,)
\ee
be the coordinates of a final event $\,F\,,$ involving the return of 
the signal to the initial point after its reflection. Represent 
the components of the coordinate velocity of the reflected signal 
in the form 
\be \lb{65}
{\tilde v}_1=-\frac{dx_1}{dx_0^{(2)}}\,;\quad {\tilde v}_2=-
\frac{dx_2}{dx_0^{(2)}}\,;\quad
{\tilde v}_3=-\frac{dx_3}{dx_0^{(2)}}\,\,.
\ee
It was mentioned before that the reflected signal propagates within
the sector $\,\Psi \zeta \Omega \epsilon o\,$ conjugate to the sector 
$\,\Gamma \gamma \Delta \beta o\,.$ It is therefore clear from Fig.\,3
that $\,{\tilde v}_{\alpha }\ne -v_{\alpha }\,$ and, consequently, 
$\,dx_0^{(2)}\ne dx_0^{(1)}\,.$ In
virtue of (65), (63), (59), we have 
$$
\begin{array}{rl}
-v_1\,/\,\tilde v_1=-v_2\,/\,\tilde v_2=-v_3\,/\,\tilde
v_3=&\,\\ 
dx_0^{(2)}\,/\,dx_0^{(1)}=v_2+v_3-v_1&.\\
\end{array}
$$
The latter equality, together with (60), makes up a system of two equations.
It will be written as
$$
\left\{
\begin{array}{lcl}
dx_0^{(2)}\,/\,dx_0^{(1)}&=&2(v_2+v_3)-1\\
dx_0^{(2)}\,/\,dx_0^{(1)}&=&1-2v_1\,.\\
\end{array}
\right.
$$ 
Hence, taking into account (63), we obtain the
following relations 
\be \lb{66}
(\,dx_0^{(1)}+dx_0^{(2)}\,)\,/\,2=dx_2+dx_3\,,
\ee
\be \lb{67}
(\,dx_0^{(1)}-dx_0^{(2)}\,)\,/\,2=dx_1\,.
\ee

Turning to the definition (64), it is easily understood that the quantity
$\,(\,dx_0^{(1)}+dx_0^{(2)}\,)\,/\,2\,$ prescribes the 3D distance $\,dl\,$ 
between the events $\,I\,$ and $\,R\,.$ By definition, these 
events have 3D coordinates $\,(\,0\,,0\,,0\,)\,$ and 
$\,(\,dx_1\,,dx_2\,,dx_3\,)\,,$ respectively, in which case
the vector $\,d\mbf x=(\,dx_1\,,dx_2\,,dx_3\,)\,$ falls into the sector 
$\,\Gamma \gamma \Delta \beta o\,.$ Thus, within the
given sector, the relation (66) gives 
\be \lb{68}
dl=dx_2+dx_3\,.
\ee

We now consider a procedure which allows synchronization of coordinate
clocks (i.e.\ clocks reading the coordinate time $\,x_0\,$) located at the
neighbouring points $\,\cal I\,$ and $\,\cal R\,$ of 3D space; we intend to
determine the difference $\,\Delta x_0\,$ between the readings of
these neighbouring clocks, which corresponds to simultaneous  events at $\,\cal
I\,$ and $\,\cal R\,.$ 

Let $\,(\,0\,,0\,,0\,)\,$ and $\,(\,dx_1\,,dx_2\,,dx_3\,)$ be the 3D coordinates
of 
the points $\,\cal I\,$ and $\,\cal R\,,$ respectively. 
Choose as one of the events the event $\,R\,$ at
the point $\,\cal R\,$ which has the coordinates
$\,(\,dx_0^{(1)}\,,dx_1\,,dx_2\,,dx_3\,)\,.$
Then another event $\,S\,$ at the point $\,\cal I\,,$ with coordinates 
$\,(\,(dx_0^{(1)}+dx_0^{(2)})\,/\,2\,,0\,,0\,,0\,)\,$ is obviously 
simultaneous to the event
$\,R\,$ at the point $\,\cal R\,.$ As a result 
$$
\Delta x_0=dx_0^{(1)}-\frac{dx_0^{(1)}+dx_0^{(2)}}{2}=
\frac{dx_0^{(1)}-dx_0^{(2)}}{2}\,.
$$ 
Using the relation (67), we finally find that 
\be \lb{69}
\Delta x_0=dx_1\,.
\ee
This formula permits synchronization of clocks within the sector 
$\,\Gamma \gamma \Delta\beta o\,.$ 

Moreover, consider the motion of a particle and determine $\,v\,,$ i.e.\ the
observable values of its velocity. For obtaining $\,v\,,$ it is necessary
first to know the true time $\,d\tau \,,$ spent by this particle on the
displacement $d\mbf x=(\,dx_1\,,dx_2\,,dx_3\,)\,.$ 

If the particle starts from point $\,\cal I\,$  at an instant of
coordinate time $\,0\,$ and reaches point $\,\cal R\,$ at an instant
of coordinate time $\,dx_0\,$, then the true time $\,d\tau \,$ spent
on the displacement is not equal to
$\,dx_0\,$ but equal to the difference between the instants $\,dx_0\,$ and 
$\,\Delta x_0\,$ which is simultaneous at $\,\cal R\,$ to the instant $\,0\,$ 
at the starting point $\,\cal I\,,$ i.e.\ $\,d\tau =dx_0-\Delta
x_0\,.$ Thus, using (69), 
we get 
\be \lb{70}
d\tau =dx_0-dx_1\,.
\ee
As a result, from (68) and (70)  
\be \lb{71}
v=\frac{dl}{d\tau }=\frac{dx_2+dx_3}{dx_0-dx_1}=\frac{v_2+v_3}{1-v_1}\,.
\ee
The formula given shows how within the sector 
$\,\Gamma \gamma \Delta \beta o\,$  the observable value of the
particle velocity is expressed in terms of the components 
$\,v_1\,,v_2\,,v_3\,$ of its coordinate velocity. According to (71), 
$\,v\le 1\,,$ with  $\,v=1\,$ for a photon. In the latter case, (71)
is equivalent to (54) and thus to the light cone equation $\,ds=0\,.$ 

After similar calculations for each of the remaining eleven sectors we
obtain the complete set of formulae which determine the observables. These
formulae are tabulated in the Table of the observables. According to
this Table, the symmetry of 3D space is determined not by the rotation
group but by a corresponding group of discrete transformations: the flat
3D space which corresponds to the totally anisotropic event space (43) is
non-Euclidean. This is demonstrated most easily if we graphically 
reproduce an Euclidean image of the sphere of radius $dl\,$
prescribed in the flat non-Euclidean 3D space. For this purpose, a 
rectangular system of coordinates $\,dx_1\,,dx_2\,,dx_3\,$ is
introduced in Euclidean 3D space and use is made of
the relations presented in the second column of the Table. It can readily be
seen that each of the twelve sectors cuts its own piece (a rhomb) out of the
corresponding plane $\,dl=const\,.$ All twelve rhombs turn out to be
equal to each
other and taken together constitute the surface of a regular rhombic
dodecahedron. Such a dodecahedron is illustrated in Fig.\,4. The Cartesian
coordinates of 14 vertices of the dodecahedron are represented as rectangular
components of the corresponding radius vector 
$$
\begin{array}{lll}
\overleftarrow{\alpha o}=dl(\,1\,,0\,,0\,)\,;\quad &\overleftarrow{\beta o
}=dl(\,0\,,1\,,0\,)\,;\quad &\overleftarrow{\gamma o}=dl(\,0\,,0\,,1\,)\,;\\
\overleftarrow{\delta o}=dl(\,-1\,,0\,,0\,)\,;\quad &\overleftarrow{
\epsilon o}=dl(\,0\,,-1\,,0\,)\,;\quad &\overleftarrow{\zeta o
}=dl(\,0\,,0\,,-1\,)\,;\\
\end{array}
$$
$$
\begin{array}{ll}
\overleftarrow{\Gamma o}=dl(\,1/2\,,1/2\,,1/2\,)\,;\quad
&\overleftarrow{\Delta o}=dl(\,-1/2\,,1/2\,,1/2\,)\,;\\
\overleftarrow{\Theta o}=dl(\,-1/2\,,-1/2\,,1/2\,)\,;\quad
&\overleftarrow{\Lambda o}=dl(\,1/2\,,-1/2\,,1/2\,)\,;\\
\overleftarrow{\Xi o}=dl(\,1/2\,,1/2\,,-1/2\,)\,;\quad
&\overleftarrow{\Phi o}=dl(\,-1/2\,,1/2\,,-1/2\,)\,;\\
\overleftarrow{\Psi o}=dl(\,-1/2\,,-1/2\,,-1/2\,)\,;\quad
&\overleftarrow{\Omega o}=dl(\,1/2\,,-1/2\,,-1/2\,)\,.\\
\end{array}
$$
By the coordinates of the vertices it is easy to calculate an acute angle of any
rhomb, e.g.\ $\angle{\,\Gamma\gamma\Delta}\,.$ It turns out that 
$$
\angle{\,\Gamma\gamma\Delta}=\arccos\frac{1}{3}\approx 70^\circ\,.
$$

Needless to say that in comparison with Minkowski space
the relativistically invariant Finslerian space-time (43) -with entirely
broken isotropy of 3D space- possesses more exotic properties than the
relativistically invariant Finslerian space-time (6) with partially
broken isotropy. In spite of the fact that, proceeding from the
flat metric (43), it is easy to build the corresponding model of a curved
Finslerian space possessing local relativistic symmetry and local entire
3D anisotropy, it is still diffucult to indicate the place which such a model
could occupy in modern physics.\footnote{A possible speculation would
  be that the flat Finslerian metric (43) describe the space geometry in the
asymptotically free limit of quantum chromodynamics, i.e.\ at distances much
smaller than the tenth part of a fermi.} 

\bigskip
\bigskip

\noindent
{\bf 6. A FIELD OF LOCAL ANISOTROPY AND\\
\phantom{Aa}THE FINSLERIAN MODEL OF A CURVED SPACE-TIME}

\bigskip
\bigskip

It is obvious that within the framework of the model of  flat
Finslerian spaces as given by (6) or (43), it is impossible to answer 
constructively the question of the possible origin of
local anisotropy. While discussing the physical nature of inertia,
Mach arrived at the conclusion that it is unreasonable to speak of the
acceleration of a body relative to empty space. Inertia of bodies
should be regarded as their ability to resist acceleration relative to
external matter. Since external matter is distributed
nonuniformly, inertia and inertial forces arising from acceleration should
depend on the localization of a body and on the direction of its acceleration.
Consequently, inertial mass should be a quantity represented
by a {\em tensor} field over space-time. When this conclusion is
compared with the fact that inertial mass in anisotropic space is
represented by a tensor, such a comparison suggests that the
parameters $r$ and $\nub \,,$ in terms of which the inertial mass
(29) is expressed, should be regarded not as constants but as fields over
space-time with a matter distribution as their source. Consequently,
we should also consider a space-time with local
anisotropy varying from point to point. Then, due to the dependence on the
fields $r$ and $\nub $ characterizing the local anisotropy of what
will turn out to be a curved space-time, the inertial mass (29) will
acquire the character of a tensor field in correspondence with Mach's 
principle. In relativistic metric theories of gravitation, where
$r=0\,,$ such a result cannot be obtained. 

The Finslerian metric of a curved locally anisotropic space-time must be of
such a form that, on the one side, the principle of correspondence
with the Riemannian metric of a  curved locally isotropic
space-time of GR is satisfied, and on the other side, at any point it
ought to admit a representation in the form (6)\footnote{or (43)}. The
Finslerian metric with the above-mentioned properties turns out to be the
following 
\be \lb{72}
ds = \left [\frac{(\,\nu _i\,dx^i\,)^2}{g_{ik}\,dx^idx^k}\right ]^{r/2}\,
\sqrt{g_{ik}\,dx^idx^k}\,.
\ee
The given metric is a function of three fields: $r=r\,(\,x\,)$, a
scalar field determining the magnitude of local anisotropy; $\nu
_i=\nu _i\,(\,x\,)$, a vector field of locally preferred directions in
space-time satisfying the condition $\nu _i\,\nu ^i =
g_{ik}\,\nu ^i\,\nu ^k = 0\,,$ and finally $g_{ik} =
g_{ik}\,(\,x\,)\,,$ the field of a Riemannian metric tensor. At each
of its points, the curved Finslerian space-time (72) has its own
tangent space (6) with its own values of the parameters $r$ and $\nub $
which determine the local anisotropy. These values of the parameters are none
other than the local values of the corresponding fields $r\,(\,x\,)$ and $\nu
_i\,(\,x\,)\,.$ 

The metric (72) is written in arbitrary coordinates. It is therefore important
to elucidate how the difference of the coordinates of two neighbouring events
is related to observables. First of all consider proper time. From
(72) the interval $d\tau \,,$ measured by an observer at rest at a point
with spatial coordinates $x^\alpha \,,$ is related to the interval $dx^0$ of
coordinate time by the relation $c\,d\tau = \left (\nu _0^2\,/\,g_{00}
\right )^{r/2}\,\sqrt{g_{00}}\,dx^0\,.$
For obtaining the 3D distance between neighbouring points and for
synchronizing the coordinate clocks it is necessary to use the
exchange of light signals. This can easily be done since the light
cone equation remains the same as in GR (in accordance with (72)). As a
result, the 3D metric turns out to be the following: $dl^2=\gamma _{\alpha \beta
}\,dx^\alpha dx^\beta \,,$   
where $\gamma _{\alpha \beta}= \left (\nu _0^2\,/\,g_{00}\right )^r
\left (-g_{\alpha \beta}+g_{0\alpha }g_{0\beta }/g_{00}\right )\,$ and
the difference $ \Delta x^0 $ of the readings of the coordinate clocks
recording the simultaneous events at the neighbouring points is given by the
formula $ \Delta x^0=-g_{0\alpha }\,dx^{\alpha} \,/\,g_{00}\,.$ 

The structure of the locally anisotropic Finslerian space (72) is such that the
motions of massless particles and of test bodies in it are significantly
different. Light propagates along Riemannian geodesics with the
metric tensor $g_{ik}$ whereas free fall of test bodies occurs along Finslerian
geodesics [33]. 

According to (72), the dynamics of Finslerian space-time is completely
determined by the dynamics of the gravitational field
$g_{ik}\,(\,x\,)\,$ and of the fields $r\,(\,x\,)\,$
 and $\nu _i\,(\,x\,)\,,$ responsible for local anisotropy. Since
 these three fields interact with each other and with matter, for a
 description of the dynamics it is necessary to
 construct equations which generalize the corresponding Einstein
 equations. The key role in solving this task is played by the
 property of invariance of the Finslerian metric (72) under the
transformations 
\be \lb{73}
g_{ik}\ \to\ e^{2\sigma (\,x\,)}\,g_{ik}\,,\qquad 
\nu _i\ \to\ e^{(\,r-1\,)\sigma (\,x\,)\,/\,r}\,\nu _i\,,
\ee
where $\sigma (\,x\,)$ is an arbitrary function. Apart from the metric, the
local
transformations (73) leave invariant all the observables. Therefore in the
theory taking account of the anisotropy of space-time the transformations (73)
are local gauge transformations. Gauge-invariant, for example, is the action
for a  compressible fluid in the Finslerian space [34]
$$S=-\frac{1}{c}\int \mu ^\ast
\left(\frac{\nu_i\,v^i}{\sqrt{\,g_{ik}\,v^i\,v^k}}\right )^{4r}\sqrt{-g}
\,\,d^{\,4}x\,,$$
where $\,\mu ^\ast\,$ is the invariant fluid energy density, $v^i=dx^i/ds\,,$
and $ds$ is
the Finslerian metric (72). 

In connection with the mentioned gauge invariance, the dynamic system
consisting of the fields $\,g_{ik}\,, r\,, \nu _i\,$ and a
compressible fluid must be complemented with a vector gauge field
$\,B_i\,$ which under (73) transforms as follows
$$B_i\ \to\ B_i+b\,[\,(\,r-1\,)\,\sigma (\,x\,)\,/\,r\,]_{;\,i}\ ,$$
where $\,b\,$ is a constant with a dimension of length. As a result,
the behaviour of the given system is described by the following gauge-invariant
variational
principle
$$
\delta \int\left\{-\frac{1}{2}[\cdots ]R-\frac{3}{4}
[\cdots ]^{-1}[\cdots ]^{;\,i}[\cdots ]_{;\,i}-\frac{r^{;\,i}r_{;\,i}}
{4\varsigma (\varepsilon -r)}\left (\frac{\cdots }{\cdots }\right )^{\,2r}
\right.
$$
$$
-\frac{f}{4}{\cal N}_{ik}{\cal N}^{ik}\left (\frac{\cdots }{\cdots }
\right )^{2r-2}
+\frac{1}{2}\lambda ^2f\nu _i\nu ^i\left (\frac{\cdots }{\cdots }
\right )^
{4r-2}-\frac{1}{4}{\cal F}_{ik}{\cal F}^{ik}
$$
\be \lb{74}
\left.-\frac{8\pi \hat k}{c^4}
\mu ^\ast
\left(\frac{\nu_i\,v^i}{\sqrt{\,g_{ik}\,v^i\,v^k}}\right )^{4r}\right\}
\sqrt{-g}
\,\,d^{\,4}x=0\,,
\ee
where\,\ \ $\left (\frac{\cdots }{\cdots }\right )=\left (\nu ^kr_{;\,k}/
\sqrt{-r_{;\,k}r^{;\,k}}\right )\,,\ \ [\cdots ]=[(1-r/\varepsilon )
\left (\frac{\cdots }{\cdots }\right )^{\,2r}]\,,\ \ {\cal N}_{ik}=
\nu _{k;\,i}-\nu _{i;\,k}-(\nu _kB_i-\nu _iB_k)/b\,,\ \ {\cal F}_{ik}=
B_{k;\,i}-B_{i;\,k}\,,\ R$ is a Riemannian scalar. The constants 
$f\,,\,\varsigma $ and $1/\varepsilon $ are dimensionless; $\varsigma
$ characterizes the interaction of the fluid
(matter) with the field $r\,$ while $1/\varepsilon \,,$ the interaction of the
fields $r$ and $g_{ik}\,;$ $\hat k$ is a gravitational constant related
to the observable Newtonian constant by $\hat k=k/\eta \,;\,\,$ $\eta$
is a renormalization constant given by the formula
$$
\eta =1+\frac{\varsigma /(2\varepsilon
)}{[1+\varsigma/(4\varepsilon)]^{1/2}} 
$$
and, finally, $\lambda ^2$ is a Lagrange multiplier. 

The variational principle (74) leads to the equations of relativistic
hydrodynamics in the locally anisotropic space and also to a system of
gauge-invariant field equations. In a gauge given by the condition  $\nu
^kr_{;\,k}=\sqrt{-r_{;\,k}r^{;\,k}}\,,$
the corresponding system of field equations is presented
in [12, 13]. It should be noted that if the existence of a ``fifth force'' is
confirmed then the gauge field $B_i$ may be regarded as its carrier. An
additional term $\sim \,B_ij^i\,$ must then be incorporated in the variational
principle (74), where $j^i$ is a preserved current involved in the
hydrodynamic equations\footnote{Such a refinement of the variational principle
(74) actually seems to be necessary. In this connection see the report [35].}. 

In ref.\,[11], the static centrally symmetric solution of the new field
equations was found, i.e.\ the Finslerian problem of Schwarzschild
solved. Subsequently, in a post-Newtonian approximation the equations
of Finslerian geodesics were
integrated and corrections to the classical gravitational effects arising from
the local anisotropy of space-time were calculated. Comparison of these
corrections with error estimates in the experimental data relating to the
solar system gives the following constraints on the interaction constants 
$$
- 0,054 < \varsigma \le 0\,\,,\qquad 0 < 1/\varepsilon < 0,25\,.
$$

Within the framework of the Finslerian theory, the equality $\varsigma
=0$ means the
absence of the field $r$ determining the magnitude of the local
anisotropy of space-time. In this case the Finslerian metric (72) reduces to
the Riemannian one and the Finslerian gravitation theory to the Einstein
theory. If $\,\varsigma \ne 0\,$, the presence of the field of locally
preferred directions introduces a partial ordering into the structure
of space-time; it is precisely by this that the Finslerian  space-time
(72) is distinguished from the ``amorphous'' Riemannian space. It must
be added here that, according to
the field equations of the Finslerian theory, the main source of the field $r$
is the trace of the energy-momentum tensor for the matter fields which is
zero for the massless and nonzero for massive fields. As a result,
a scenario of the evolution of the Universe becomes possible where only
initially, i.e.\  before the appearance of high-temperature phase
transitions with a successive breaking of higher gauge symmetries and
before the appearance of masses in the fundamental matter field ,
space-time was Riemannian. With the appearance of massive elementary 
particles, the trace of the energy-momentum tensor becomes nonzero. In
this case, a strong local anisotropy of space-time is generated, i.e.\
there occur phase transitions in its local geometric structure as a
result of which space-time acquires a Finslerian metric. In the course
of the subsequent expansion the initially strong local anisotropy of
space gradually decreases and on the average tends to zero along with
its curvature. Thereby, the Finslerian space-time again approximates a
Riemannian one. Apparently, it is the induced phase transitions in the
geometric structure of space-time which make energetically most
favourable the scheme of breaking higher gauge symmetries realized in nature. 

\bigskip
\bigskip
\bigskip

\noindent
{\bf 7. CONCLUSION}

\bigskip
\bigskip 

We descibed two types of Finslerian event spaces, namely, spaces with 
partially and entirely broken local rotational symmetry in 3D
space. Since the locally isotropic Riemannian space-time is a special
case of the Finslerian space-time (72) (corresponding to the vanishing
of the field $\,r\,$), one
can speak of a joint description of three geometric models of
space-time. It is important to
stress that each of the above-mentioned models possesses (differing) local
relativistic invariances. Depending on the magnitude and character of
the breaking of local 3D isotropy, local relativistic invariance may
take either the form of full Lorentz invariance (3D rotational
symmetry not broken), the form of generalized Lorentz invariance, i.e.
invariance under the transformations (7) (partial breaking of
isotropy), or invariance under the transformations (45) (total
breaking of isotropy). 

Experimental discoveries of recent years, in particular the
discovery of the anisotropy of the cosmic background radiation have
led to a renaissance of interest in theories with a preferred frame of
reference. In essence, in such investigations, the old idea of an
absolute ``ether'' is exploited, the only difference
being that the preferred frame is now identified with a frame in which
the cosmic background radiation is locally isotropic and the already 
established physical laws are operative. In this case, attempts are
sometimes undertaken to explain new experimental results by an {\em ad-hoc}
breaking of Lorentz invariance in the passage from the preferred frame to
another (laboratory) inertial frame. In this way, certainly ``anything''
can be ``explained''. At the same time, Einstein's principle of
relativity, implemented with help of the generalized Lorentz
transformations, allows the avoidance of such a diversity of options,
and convincingly leads to local anisotropy of space-time. As a result,
the problem of a possible violation of the Lorentz transformations
reduces to the problem of existence of local anisotropy of
space-time.\footnote{A different approach to deviations from Lorentz 
invariance which, however, leads to a more complicated physics was
followed in refs. [36].}
 
In connection with a possible local anisotropy of space-time it will
be recalled
that according to the model of the hot Universe the temperature of relic
radiation should not depend on the direction in which it is being measured. At
the same time the temperature anisotropy of relic radiation is already an
experimental fact with dipole component of anisotropy having the
largest value. Investigators usually do not express a fundamental interest
in such a dipole anisotropy because they believe that it arises from the fact
that our lab frame accidentally moves at a certain velocity
relative to the cosmic microwave background. Such an explanation would
be  more satisfactory if the corresponding anisotropy were also
observed in the Hubble constant. Until now, studies of the
angular dependence of the Hubble constant are neither precise enough
nor covering a larger section of the sky. ( Cf. [37].) If a special analysis
will show that there
is no correlated dipole anisotropy in the Hubble
constant then the dipole anisotropy of relic radiation might be an
indication of a strong local anisotropy of space-time at an early stage of the
evolution of the Universe. The point is that in a space with strong anisotropy
there indeed
exists a physically preferred frame; with respect to this frame the
hot background radiation was isotropic while the velocity distribution
of massive relativistic particles was anisotropic. As a result, the
Hubble constant became anisotropic. Therefore, by passage to
another frame, a reversed situation becomes possible: the Hubble
constant looses its dipole anisotropy while the background radiation
picks it up.

It has already been noted that the experimental data on the behaviour of the
spectra of primary ultra-high energy cosmic protons were one of the motivations
for the Finslerian generalization of relativity theory. In spite of indirect
evidence in
favour of it, the relativistic theory of locally anisotropic space-time,
outlined in the present paper, is still in need of empirical
support. Since the alternative to local anisotropy is a
strict local isotropy of space-time, and since in nature any strict symmetry
holds only approximately, it seems reasonable to continue
investigations into the physical manifestations of local anisotropy. 
In fact, such a line of research is equivalent to the testing of
SR and of Lorentz invariance to which increased attention has been
paid recently, both from the experimental [38] and theoretical side [39].\\

\newpage
{\bf REFERENCES}

\bigskip
\bigskip

\begin{itemize} 
\item[1.] Rund, H. (1959). {\em The Differential Geometry of Finsler
Spaces} (Springer, Berlin). 
\item[2.] Doi, T., et al. (1995). In {\em Proc. 24th Int. Cosmic Ray Conf.},
{\bf 2} (Roma),740.
\item[3.] Greisen, K. (1966). {\em Phys. Rev. Lett.} {\bf 16}, 748.
\item[4.] Zatsepin, G. T., and Kuz'min, V. A. (1966). {\em Pis'ma Zh. Eksper.
Teor. Fiz.} {\bf 4}, 114. 
\item[5.] Kirzhnits, D. A., and Chechin, V. A. (1971). {\em Pis'ma Zh. Eksper.
Teor. Fiz.}\\{\bf 14}, 261.
\item[6.] Kirzhnits, D. A., and Chechin, V. A. (1972). {\em Yadern. Fiz.} {\bf
15}, 1051.
\item[7.] Khristiansen, G. B. (1974). {\em Cosmic Rays of Superhigh Energies}\\ 
(Moscow Univ. Press, Moscow), in Russian.
\item[8.] Coleman, S., and Glashow, S. L. (1998). Preprint hep-ph/9808446.
\item[9.] Berezinsky, V. S. (1998). Preprint INFN/TH-98/03.
\item[10.] Bogoslovsky, G. Yu. (1977). {\em Nuovo Cimento B}{\bf 40}, 99.
\item[11.] Bogoslovsky, G. Yu. (1992). {\em Theory of Locally Anisotropic
Space-Time}\\
(Moscow Univ. Press, Moscow), in Russian.
\item[12.] Bogoslovsky, G. Yu. (1992). {\em Class. Quantum Grav.} {\bf 9}, 569.
\item[13.] Bogoslovsky, G. Yu. (1994). {\em Fortschr. Phys.} {\bf 42},
143; (1993).
{\em Phys. Part. Nucl.} {\bf 24}, 354.
\item[14.] Cunningham, E. (1910). {\em Proc. London Math. Soc.} {\bf 8}, 77.
\item[15.] Bateman, H. (1910). {\em Proc. London Math. Soc.} {\bf 8}, 223.
\item[16.] Fulton, T., Rohrlich, F., and Witten, L. (1962). {\em Rev. Mod.
Phys.}
{\bf 34}, 442.
\item[17.] Bogoslovsky, G. Yu., and Goenner, H. F. (1995). In {\em Abstr. Int.
School-Seminar 
``Found. of Grav. and Cosmology"}, (Odessa), 79.
\item[18.] Bogoslovsky, G. Yu., and Goenner, H. F. (1997). \ In {\em Abstr. 5th
Int.
Wigner\\Symposium}, (Vienna), 20.
\item[19.] Bogoslovsky, G. Yu., and Goenner, H. F. (1998). {\em Phys. Lett.
A}{\bf 244}, 222.
\item[20.] Riemann, B. (1867). {\em Abhandlungen der K\"onigl. Gesellschaft
  d. Wissenschaften,\\G\"ottingen} {\bf 13}, 14.
\item[21.] Winternitz, P., and Fri\v{s}, I. (1965). {\em Yadern. Fiz.} {\bf 1},
889.
\item[22.] Goenner, H. F., and Bogoslovsky, G. Yu. (1997). Preprint
gr-qc/9701067;\\ 
Submitted to {\em Gen. Rel. Grav.}
\item[23.] Tavakol, R. K., and Van den Bergh, N. (1985). {\em Phys. Lett. A}{\bf
112}, 23.
\item[24.] Tavakol, R. K., and Van den Bergh, N. (1986). {\em Gen. Rel. Grav.}
{\bf 18}, 849.
\item[25.] Bogoslovsky, G. Yu. (1976). {\em Pis'ma Zh. Eksper. Teor. Fiz.} {\bf
23}, 192.
\item[26.] M{\o}ller, C. (1962). {\em Proc. R. Soc. A}{\bf 207}, 306.
\item[27.] Champeney, D. C., Isaak, G. R., and Khan, A. M. (1963). {\em Phys.
Lett.} {\bf 7}, 241.
\item[28.] Kaivola, M., Poulsen, A., et al. (1985). {\em Phys. Rev. Lett.} {\bf
  54}, 255;\\McGowan, R. W., Giltner, D. M., et al. (1993). {\em
  Phys. Rev. Lett.} {\bf 70}, 251.
\item[29.] Isaak, G. R. (1970). {\em Phys. Bull.} {\bf 21}, 255.
\item[30.] Bogoslovsky, G. Yu. (1977). {\em Nuovo Cimento B}{\bf 40}, 116.
\item[31.] Barbour, J. B. (1974). {\em Nature}, 328; (1975). {\em Nuovo Cimento
B}{\bf 26}, 16;\\Barbour, J. B., and Bertotti, B. (1977). {\em Nuovo Cimento
B}{\bf
  38}, 1.\\Cf. the review of Goenner, H. (1981). In {\em Grundlagenprobleme der
  modernen Physik},\\Eds. Nitsch, J., et al. 
(Bibliographisches Institut, Mannheim), 85.
\item[32.] Noether, E. (1918) {\em G\"ottinger Nachrichten. Math. Phys. Kl} {\bf
H2}, 235.
\item[33.] Bogoslovsky, G. Yu. (1984). {\em Ukr. Fiz. Zh.} {\bf 29}, 17.
\item[34.] Bogoslovsky, G. Yu. (1986). {\em Dokl. Akad. Nauk SSSR} {\bf 291},
317.
\item[35.] Anderson, J. D., et al. (1998). {\em Phys. Rev. Lett.} {\bf 81},
2858.
\item[36.] Bleyer, U., and Liebscher, D. E. (1986). {\em Astronomische
  Nachrichten} {\bf 307}, 267;\\Bleyer, U. (1988). In {\em
  Gravitation und Kosmos}, Ed. Wahsner, R.\\(Akademie-verlag, Berlin), 91.
\item[37.] Ichikawa, T., and Fukugita, M. (1992). {\em Astrophysical Journal}
{\bf
  394}, 66.
\item[38.] Nielsen, H. B., and Picek, I. (1982). {\em Phys. Lett. B}{\bf 114},
141;\\
    Fischbach, E., Haugan, M. P., Tadic, D., and Hai-Yang-Cheng. (1985).
    {\em Phys. Rev. D}{\bf 32}, 154.   
\item[39.] Mansouri, R., and Sexl, R. U. (1977). {\em Gen. Rel. Grav.} {\bf 8},
497; 515;\\Golestanian, R., Khajehpour, M. R. H., and Mansouri,
R. (1995). {\it Class. Quantum Grav.} {\bf 12}, 273.

\end{itemize}
\newpage

\vspace*{3cm}

\noindent
{\bf TABLE AND FIGURE CAPTIONS}\\
 
\noindent
Tabl.\,I : Table of observables.\\

\noindent
Fig.\,1 : Plots for $d\tau /dt=\left [(1-v\cos\alpha /c)/\sqrt{1-v^2/c^2}\right
] ^r\sqrt
{1-v^2/c^2}$ at $\,r=0.6\,$ and at three successive values $\,0\,,\,\pi
/4\,,\,\pi /2\,$
of the angle $\,\alpha \,$ between $\,\mbf{v}\,$ and $\,\nub\,.$ These plots 
demonstrate the specific features of the behaviour of the anisotropic factor of 
time dilatation $(d\tau /dt)\!\!\!\mid_{r>0}$ in comparison with the behaviour
of 
the isotropic
(Minkowskian) factor $(d\tau /dt)\!\!\mid_{r=0}\,.$ \\

\noindent
Fig.\,2 : Parametric 3D plots illustrating the dependence of deformation of a
two-sheet hyperboloid on the magnitude $\,r\,$ of space anisotropy. Any of the
deformed hyperboloids remains inscribed into a light cone and like a light cone
it is an invariant of the generalized Lorentz transformations (22). \\

\noindent
Fig.\,3 : The relativistically invariant range of permissible $v_\alpha\,$
values. \\

\noindent
Fig.\,4 : A regular rhombic dodecahedron as an Euclidean image of the sphere of
radius
$dl\,,$ prescribed in the flat non-Euclidean 3D space. \\

\newpage

\vspace*{5cm}

\begin{center}
Table of observables
\vskip 5mm
\begin{tabular}{|r c|c|c|c|}
\hline
%sector&  & d$l$ &  $\Delta x_0$  &  $v$ \\
\multicolumn{2}{|c}{sector}
    &\multicolumn{1}{|c}{$dl$}
      &\multicolumn{1}{|c}{$\Delta x_0$}
        &\multicolumn{1}{|c |}{$v$}\\
\hline \hline
$\Gamma \gamma\Delta \beta o$ &  & $dx_2+dx_3$
& $dx_1$ & $(v_2+v_3)/(1-v_1)$ \\
\hline
 &$\Psi \zeta \Omega \epsilon o$ & $-(dx_2+dx_3)$
& $dx_1$ &$-(v_2+v_3)/(1-v_1)$ \\
\hline 
$\Gamma \beta \Xi \alpha o$ &  & $dx_1+dx_2$
& $dx_3$ & $(v_1+v_2)/(1-v_3)$ \\
\hline
 &$\Psi \epsilon \Theta \delta o$ & $-(dx_1+dx_2)$
& $dx_3$ & $-(v_1+v_2)/(1-v_3)$ \\
\hline
$\Gamma \alpha \Lambda \gamma o$ &  & $dx_1+dx_3$
& $dx_2$ & $(v_1+v_3)/(1-v_2)$ \\
\hline
 &$\Psi \delta \Phi \zeta o$ & $-(dx_1+dx_3)$
& $dx_2$ & $-(v_1+v_3)/(1-v_2)$ \\
\hline
$\Omega \epsilon \Lambda \alpha o$ &  & $dx_1-dx_2$
& $-dx_3$ & $(v_1-v_2)/(1+v_3)$ \\
\hline
 &$\Delta \beta \Phi \delta o$ & $-(dx_1-dx_2)$
& $-dx_3$ & $-(v_1-v_2)/(1+v_3)$ \\
\hline
$\Omega \alpha \Xi \zeta o$ &  & $dx_1-dx_3$
& $-dx_2$ & $(v_1-v_3)/(1+v_2)$ \\
\hline
 &$\Delta \delta \Theta \gamma o$ & $-(dx_1-dx_3)$
& $-dx_2$ & $-(v_1-v_3)/(1+v_2)$ \\
\hline
$\Theta \gamma \Lambda \epsilon o$ &  & $dx_3-dx_2$
& $-dx_1$ & $(v_3-v_2)/(1+v_1)$ \\
\hline
 &$\Xi \zeta \Phi \beta o$ & $-(dx_3-dx_2)$
& $-dx_1$ & $-(v_3-v_2)/(1+v_1)$ \\
\hline
\end{tabular}
\end{center}

\end{document}